\documentclass[a4paper]{amsart}
\usepackage{xcolor}
\usepackage{amsmath, amsthm, amsfonts, amssymb}
\usepackage{tikz-cd} 
\usepackage{tikz-3dplot}
\usepackage[siunitx, oldvoltagedirection]{circuitikz}
\usepackage{pgfplots}
\usepackage{pgfplotstable}
\usepackage{enumerate}
\usepackage[onecolumn]{optional}
\newtheorem{thm}{Theorem}[section]
\newtheorem{cor}[thm]{Corollary}
\newtheorem{lem}[thm]{Lemma}
\newtheorem{prop}[thm]{Proposition}
\theoremstyle{definition}
\newtheorem{defn}[thm]{Definition}
\newtheorem{exmp}[thm]{Example}
\theoremstyle{remark}

\newcommand\Fi{\mathcal{F}}
\newcommand\Ge{\mathcal{G}}
\newcommand\Pa{\mathcal{Q}}
\newcommand\RE{\mathbb{R}}
\newcommand{\bscdot}{\boldsymbol{\cdot}}
\newcommand{\Zint}[1]{[1\mathrel{{.}\,{.}}\nobreak #1]}
\DeclareMathOperator{\interior}{int}
\DeclareMathOperator{\boundary}{bdr}
\DeclareMathOperator{\closure}{cl}
\DeclareMathOperator{\co}{co}
\DeclareMathOperator{\diag}{Diag}
\DeclareMathOperator{\dist}{dist}
\DeclareMathOperator{\lcp}{LCP}
\DeclareMathOperator{\lin}{lin}
\DeclareMathOperator{\ind}{ind}
\DeclareMathOperator{\pos}{pos}
\DeclareMathOperator{\sgn}{sgn}
\DeclareMathOperator{\Sgn}{Sgn}
\DeclareMathOperator{\sm}{sm}
\DeclareMathOperator{\Span}{span}
\DeclareMathOperator*{\argmin}{arg\,min}
\numberwithin{equation}{section}
\usepgfplotslibrary{patchplots}
\usepgfplotslibrary{groupplots}
\usepgfplotslibrary{fillbetween}
\pgfplotsset{compat=1.16}
\begin{document}
\title[Equivalence of LCPs: Nonsmooth Bifurcations]{Equivalence of Linear Complementarity Problems: Theory and Application to \\ Nonsmooth Bifurcations}
\author{F\'elix A. Miranda-Villatoro}
\author{Fernando Casta\~nos}
\author{Alessio Franci}
\thanks{F\'elix A. Miranda-Villatoro is with the TRIPOP Team at the Centre de Recherche Inria de l'Universit\'{e} Grenoble-Alpes, LJK, 38000, Grenoble, France (e-mail: felix.miranda-villatoro@inria.fr).}
\thanks{Fernando Casta\~{n}os is with the Departamento de Control Autom\'{a}tico, Cinvestav-IPN, 2508 Av. IPN,
07360, Mexico City, Mexico (e-mail: fernando.castanos@cinvestav.mx).}
\thanks{Alessio Franci is with the Departamento de Matem\'{a}ticas, Universidad Aut\'{o}noma de
M\'{e}xico, Circuito exterior S/N, C.U., 04510. Mexico City, Mexico (e-mail: afranci@ciencias.unam.mx).}
\keywords{
Bifurcations, linear complementarity problems, nonsmooth dynamics, piecewise linear equations.
}
\date{\today}
\begin{abstract}
Linear complementarity problems provide a powerful framework to model nonsmooth phenomena in a variety of 
real-world applications. In dynamical control systems, they appear coupled to a linear input-output system
in the form of linear complementarity systems. Mimicking the general strategy that led to the foundation of
bifurcation theory in smooth maps, we introduce a novel notion of equivalence between linear complementarity
problems that sets the basis for a theory of bifurcations in a large class of nonsmooth maps, including, but
not restricted to, steady-state bifurcations in linear complementarity systems. Our definition exploits the
rich geometry of linear complementarity problems and leads to constructive algebraic conditions for identifying
and classifying the nonsmooth singularities associated with nonsmooth bifurcations. We thoroughly illustrate
our theory on an extended applied example, the design of bistability in an electrical network, and a more
theoretical one, the identification and classification of all possible equivalence classes in two-dimensional
linear complementarity problems.
\end{abstract}
\maketitle
\section{Introduction} \label{sec:introduction}
Complementarity conditions first appeared in the context of constrained optimization, in the form of 
the Karush-Khun-Tucker conditions to which the Lagrange multipliers are subject to~\cite{boydOpt}.
Yet, complementarity conditions emerge naturally in many other modeling contexts
including artificial intelligence, economics, and engineering. 
Electrical networks with semiconductor devices~\cite{acary2011,adly2017,goeleven2017}, 
mechanical systems with unilateral constraints~\cite{brogliato1999}, price equilibrium, 
traffic network~\cite{nagurney1993}, layers of ReLU nonlinearities used in deep neural networks~\cite{goodfellow2016deep},  and portfolio selection problems~\cite{ferris1997}
can all be modeled using a formalism in which decision variables must satisfy a set of 
complementarity conditions but the equations are otherwise linear.
Linear complementarity problems (LCPs) are nonsmooth problems in which a set of 
complementarity conditions is coupled with a set of linear relations among the problem variables.
Due to the broad spectrum of applications, complementarity problems (and LCPs in 
particular) have been extensively studied, see, 
e.g.,~\cite{murty1988,isac1992,leenaerts1998,billups2000,facchinei2003,cottle}
and references therein. Complementarity problems appear in nonlinear control theory
in the form of linear complementarity systems (LCSs), which are defined by coupling a set of
complementarity conditions to an otherwise linear input--output control system, see, 
e.g.,~\cite{schaft1998,heemels2000,camlibel2002,brogliato2003}. Biological systems like gene 
and molecule regulatory networks~\cite{casey2006piecewise} and neurons~\cite{mckean1970nagumo}
can also be modeled as LCSs.
Because of their nonlinear nature, LCSs can undergo bifurcations at
which equilibria, limit cycles, or other types of steady-state solutions appear, 
disappear, or change stability. From a control-theoretical perspective, bifurcation theory is instrumental for two
complementary reasons: stabilization, by avoiding the bifurcation through control~\cite{chen2003bifurcation,krener2004control,abed1986local,abed1987local,chen2000bifurcation}, and realization of complex nonlinear behaviors, like neuro-inspired control system~\cite{castanos2017implementing,franci2014realization,franci2019sensitivity} or bio-inspired decision making~\cite{gray2018multiagent,franci2021analysis,bizyaeva2020general,franci2019model,franci2022breaking}, by exploiting the rigorously predictable nonlinear behaviors happening close to a bifurcation. Bifurcation theory is a key control-theoretical tool in all cases in which control systems must deal with behaviors beyond equilibrium stabilization. See, e.g.,~\cite[Chapter 2]{Khalil}.
Although non-smooth bifurcation problems have been studied in a variety of contexts 
(see, e.g.,~\cite{makarenkov2012dynamics,di2008bifurcations} and references therein), a general bifurcation
 theory for LCSs is entirely undeveloped.
Here, we elaborate on our preliminary results~\cite{castanos2020}
and introduce novel theoretical tools to analyze and design the possible ways in which the set of
solutions to an LCP changes as a function of control and system parameters. We propose these
tools as candidates for the development of a non-smooth bifurcation theory for LCSs.
As opposed to existing efforts in bifurcation theory for general non-smooth dynamical systems, focusing on specific, low-dimensional, examples in a piecemeal fashion, our goal is to develop a systematic theory amenable to general  results. We propose LCSs as the framework in which to develop our theory because both mathematically tractable, due to the linear complementarity structure, and general, due to the fact that any smooth nonlinear vector field can be approximated (in the uniform metric) by a piece-wise linear one and, therefore~\cite{eaves1981a}, by an LCS.
The first contribution of our paper is to show that equilibrium bifurcations (i.e., those bifurcations solely involving equilibrium points) of 
an LCS are fully characterized by the (nonsmooth)
singularities of an auxiliary LCP. The second  contribution of our paper is to introduce of a notion of equivalence between
LCPs that allows us to characterize which LCPs are \emph{LCP-stable}, i.e., those LCPs for which the geometry of the solution space is not destroyed
by perturbations. This class of LCPs is particularly relevant in a control-theoretical setting because an
associated LCS inherits this robustness. It is also crucial to study LCSs bifurcations, because problems that are not LCP-stable constitute degenerate situations akin to organizing bifurcations~\cite{golubitsky,guckenheimer} in smooth bifurcation theory. The theory is illustrated in details on practical and mathematical examples throughout the paper.
With respect to the conference version~\cite{castanos2020} this paper: {\it i)} applies LCP singularity theory to LCS bifurcation theory, {\it ii)} provides both sufficient {\it and} necessary conditions for a matrix to be stable and it does so in arbitrary dimension, {\it iii)} uses the necessary and sufficient conditions for LCP-stability to list both LCP-stable {\it and} LCP-unstable $2\times 2$ matrices, {\it iv)} introduces the notion of {\it weak convergence} and uses it for the computation of stability margins for LCPs, {\it v)} introduces the notion of one-parameter and multi-parameter bifurcation diagram for both LCPs and LCSs and computes them in various examples.
The manuscript is organized as follows: 
Section~\ref{sec:notation} introduces the notation 
used in this work. LCSs are introduced in Section~\ref{sec:lcs}. Bifurcations in LCSs 
and their connection with LCP singularities, together with the needed background on LCPs, are introduced
in Section~\ref{sec:bifurcations}.
Sections \ref{sec:equivalence} and \ref{sec:stability} constitute the main technical
body of the paper, where the notions of LCP-equivalence and LCP-stability are studied, and 
the main technical results are presented. Subsequently, the proposed framework is applied in 
Section~\ref{sec:application} to the design of bistable LCSs and in Section~\ref{sec:classification} 
to the classification of two-dimensional LCPs.
\section{Notation} \label{sec:notation}
Let $X$ be a non-empty set. The power set of $X$ is denoted as $\mathcal{P}(X)$. 
Whenever $X$ is finite, $\vert X \vert$ denotes its cardinality. 
For a set $Y \subset X$, $Y^{c}$ denotes its complement in $X$, $Y^{c} := X \setminus Y$.
The notations $\interior Y$, $\boundary Y$ and $\closure{Y}$ stand for the interior,
the boundary and the closure of $Y$, respectively.
Given an integer $n$, we use $\Zint{n}$ to denote the interval 
of all integers between $1$ and $n$. We endow all subsets of $\mathbb{N}$
with the natural order. Let $A \in \RE^{n \times m}$, $\alpha \subseteq \Zint{n}$, and
$\beta \subseteq \Zint{m}$. The submatrix 
$A_{\alpha, \beta} \in \RE^{\vert \alpha \vert \times \vert \beta \vert}$ 
is the matrix $[A_{ij}]_{i \in \alpha,j \in \beta}$. The notation 
$A_{\bscdot, \beta}$ stands for the submatrix $A_{\Zint{n}, \beta}$.
We omit the braces whenever $\beta$ is a singleton, that is, 
$A_{\bscdot, j} = A_{\bscdot, \{j\}}$. The (polyhedral) cone generated by 
$A$ is the set of all positive linear combinations 
of the columns of $A$, that is,
\begin{displaymath}
 \pos A := \Big\{ x = \sum_{j = 1}^{m} p_{j} A_{\bscdot, j} \mid 
  p_{j} \geq 0, \, j \in \Zint{m} \Big\} \;.
\end{displaymath}
The columns of $A$ are the \emph{generators} of $\pos A$.
The cone $\pos A$ is said to be \emph{pointed} if there is no non-trivial 
subspace contained in it, whereas it is said to be \emph{strictly pointed} 
if $\ker(A) \cap \RE_{+}^{m} = \{ 0 \}$, where 
\begin{displaymath}
 \RE_{+}^{m} := \{ x\in \RE^{m} \mid x_{i} \geq 0, \, i \in \Zint{m} \}
\end{displaymath}
denotes the positive orthant of $\RE^{m}$ and $\ker(A)$ 
is the null space of $A$. 
Finally, we denote the projection of $x \in \RE^{n}$ onto the positive orthant $\RE_{+}^{n}$
by $[x]^{+}$, that is,
\begin{displaymath}
	[x]^{+} := \argmin_{w \in \RE_{+}^{n}} \frac{1}{2} \Vert x - w \Vert^{2} \;.
\end{displaymath}
\section{Linear complementarity systems} \label{sec:lcs}
Let us start with a formal definition of the linear complementarity problem.
\begin{defn}
	\label{defn:lcp}
	Given a matrix $M \in \RE^{n \times n}$ and a vector $q \in \RE^{n}$, the 
   \emph{linear complementarity problem} $\lcp(M,q)$ consists in finding a vector 
	$z \in \RE^{n}$ such that
	\begin{equation} \label{eq:complementarity:condition}
		\RE_{+}^{n} \ni M z + q \perp z \in \RE_{+}^{n} \;.
	\end{equation}
\end{defn}
Condition~\eqref{eq:complementarity:condition} is called the \emph{complementarity condition}, it 
is the compact form of the following three conditions: 1) $M z + q \in \RE_{+}^{n}$, 2) $z \in \RE_{+}^{n}$,
and 3) $(M z + q)^{\top} z = 0$.
An LCS is a linear system coupled to an LCP, i.e., a system of the form
\begin{subequations}\label{eq:Sigma}
\begin{align}
 \dot{\xi}(t) &= A \xi(t) + B z(t) + E_{1} r \label{eq:lcs:1} \;,\\
         w(t) &= C \xi(t) + D z(t) + E_{2} s \label{eq:lcs:2} \;, \\
 & \RE_{+}^{m} \ni w(t) \perp z(t) \in \RE_{+}^{m} \;, 
 \label{eq:lcs:3} 
\end{align}
\end{subequations}
where $\xi(t) \in \RE^{n}$ is the state of the system at time $t$,
$z(t)$, $w(t) \in \RE^{m}$ are the so-called complementary variables (they can 
be interpreted as external port variables), and $r \in \RE^{l}$, $s \in \RE^{l}$ are 
vectors of parameters. The matrices $A, B, C, D, E_{1}$ and $E_2$ are constant and of
appropriate dimensions. 
It is noteworthy that, even though the right-hand sides of~\eqref{eq:lcs:1}-\eqref{eq:lcs:2} are 
affine functions of the state, the time evolution of the state in~\eqref{eq:Sigma} is, 
in general, nonlinear 
(and nonsmooth) due to the complementarity condition \eqref{eq:lcs:3}.
In practice, complementarity systems such as~\eqref{eq:Sigma} arise naturally in different fields 
such as contact mechanics~\cite{brogliato1999}, electronics~\cite{acary2011, adly2017}, 
mathematical biology~\cite{acary2014}, finance~\cite{ferris1997}, etc. Moreover, in
the context of control systems,
Lur'e systems with piecewise-linear or relay-type feedback also can be written within 
the formalism of LCSs.
As an illustration, let us consider the piecewise-linear chaotic R\"ossler-like circuit~\cite{carroll1995}
\begin{align*}
	\dot{\xi}(t) & = A \xi(t) + B g(y(t)) \\
	y(t) & = C \xi(t) \\
	g(y(t)) & = 
	\begin{cases}
		             0 & \text{if $y(t) \leq 3$} \\
		\mu (y(t) - 3) & \text{if $y(t) > 3$}
	\end{cases} \;, 
\end{align*}
which can be written using complementary variables as,
\begin{align*}
	\dot{\xi}(t) & = A \xi(t) + B z(t)
	\\
	w(t) & = - \mu C \xi(t) + z(t) + 3 \mu
	\\
	& 0 \leq w(t) \perp z(t) \geq 0 \;.
\end{align*}
The fact that the piecewise-linear function $g$ above can be written in the complementarity 
formalism is not a coincidence. Indeed, there is a close relationship 
between piecewise-linear maps and linear complementarity problems, so that (under mild 
regularity assumptions) any continuous piecewise-linear map can be written as the 
solution of an LCP~\cite{eaves1981a, eaves1981b}. This property, 
together with the fact that the set of piecewise-linear functions is dense in the set of 
continuous functions with compact domain, gives LCSs a powerful framework for
approximating complicated nonlinear phenomena.
Besides continuous piecewise-linear feedback laws,
it is also possible to model some set-valued maps. Consider, for instance, the LCS
\begin{subequations} \label{eq:lure:sign}
\begin{align}
	\dot{\xi}(t) &= A \xi(t) + B 
	 \begin{bmatrix}
	 	1 & -1
	 \end{bmatrix} 
    z(t) \label{eq:lure:sign:1} \\
	w(t) &= 
	\begin{bmatrix}
		1 & 1
		\\
		1 & 1
	\end{bmatrix} z(t) + 
	\begin{bmatrix}
		1 \\ -1
	\end{bmatrix} C \xi(t) + 
	\begin{bmatrix}
		-1 \\ -1
	\end{bmatrix} \label{eq:lure:sign:2} \\
	\RE_{+}^{2} & \ni w(t) \perp z(t) \in \RE_{+}^{2} \label{eq:lure:sign:3} \;.
\end{align}
\end{subequations}
The LCP~\eqref{eq:lure:sign:2}-\eqref{eq:lure:sign:3} has the 
solution
\begin{displaymath}
 z(t) = 
  \begin{bmatrix}
    1 - C \xi(t) \\
    0	
  \end{bmatrix} \quad \text{if $C \xi(t) < 0$} \;, \quad
 z(t) = 
  \begin{bmatrix}
   0 \\
   1 + C \xi(t)
  \end{bmatrix} \quad \text{if $C \xi(t) > 0$} \;,
\end{displaymath}
and
\begin{displaymath}
 z(t) \in 
		\left\{
		\begin{bmatrix}
			1 + \theta \\
			-\theta
		\end{bmatrix}, \; \theta \in [-1, 0] \right\} \;, \quad \text{if $C \xi(t) = 0$}. 
\end{displaymath}
Hence,
\begin{displaymath}
	\begin{bmatrix}
		1 & -1
	\end{bmatrix} z(t) \in -C \xi(t) - \Sgn(C \xi(t)) \;,
\end{displaymath}
where $\Sgn: \RE \rightrightarrows [-1, 1]$ is the set-valued sign map
\begin{displaymath}
	\Sgn(x) = 
	\begin{cases}
		     -1 & \text{if $x < 0$} \\
		[-1, 1] & \text{if $x = 0$} \\
		      1 & \text{if $x > 0$}
	\end{cases} \;.
\end{displaymath}
Therefore, Eq.~\eqref{eq:lure:sign} is just another representation of a linear system with 
set-valued feedback, as those studied in set-valued sliding-mode
control~\cite{miranda2018,brogliato2020}.
Since the complementarity condition~\eqref{eq:lcs:3} is equivalent to the
inclusion
\begin{displaymath}
 -z(t) \in \mathbf{N}_{\RE_{+}^{m}}(w(t))
\end{displaymath}
with $\mathbf{N}_{S}: S \rightrightarrows \RE^{n}$ the normal cone to the set 
$S$~\cite[Proposition 1.1.3]{facchinei2003},
the LCS~\eqref{eq:Sigma} encompasses systems with state constraints~\cite{papageorgiou1989}.
In particular, when $D = 0$, Eq.~\eqref{eq:Sigma} becomes
\begin{equation} \label{eq:di:constrained}
	\dot{\xi}(t) \in A \xi(t) - B \mathbf{N}_{\RE_{+}^{m}}(C \xi(t) + E_{2} s) + E_{1} r\;.
\end{equation}
For the system to be well-defined, the trajectories of~\eqref{eq:di:constrained} must
satisfy the state constraint $C \xi(t) + E_{2} s \in \RE_{+}^{m}$ at all times $t$.
An important difference between the differential inclusions~\eqref{eq:di:constrained} and~\eqref{eq:lure:sign}
is that the former is not of Filippov type, as the normal cone $\mathbf{N}_{S}$ is not bounded at points
on $\boundary S$.
We see that the set of possible behaviors of an LCS is broad enough to model a variety of 
nonlinear phenomena involving multiple equilibria, limit cycles, sliding modes, state constraints, and chaotic behavior, to mention a few. 
\subsection*{LCS model of a circuit with bipolar transistors} \label{sec:circuit_model}
\begin{figure}
	\centering
	\begin{circuitikz}[/tikz/circuitikz/bipoles/length=0.65cm, scale=0.6, american]


 \draw[color=black]

(0,0) --++(0.8,0) node[npn,  xscale = -1.5, yscale = -1.5, anchor=B, rotate=180](npn){} 
(npn.C) |- (3,2) to [R, l =$R_{2}^{b}$] ++(0,-4) -| (npn.E)
(1.775,2) to[short, *-*]++(-2,0) to[R, l = $R_{0}^{b}$, -* ] ++(0,-2)
to[R, l=$R_{1}^{b}$, -*] ++(0,-2) to[short, *-*] ++(2,0)

(-3,2) --++(-0.8,0) node[pnp, xscale = 1.5, yscale = -1.5, anchor=B, rotate=180](pnp){} 
(pnp.C) |-  (-6,0) to[R, l=$R_{2}^{a}$] ++(0,4) -| (pnp.E)
(-4.775,0) to[short, *-]++(2,0) ++(-0.225,0) to[R, l=$R_{0}^{a}$, *-*] ++(0,2) to[R, l=$R_{1}^{a}$, -*] ++(0,2) 
to[short, -*] ++(-1.775,0)++(-0.225,0)
++(2,-2) to[R, l=$R_1$]++(3,0)
++(-3,-2) to[vR, l=$R_2$]++(3,0)
++(-1.5,-2) node[below]{$P_{0}$} to[short, o-, i<=$I$] ++(1.5,0)
++(-1.5, 6) node[above]{$P_{1}$} to[short, o-, i_=$I$] ++(-1.5,0)
;

\draw[]
(-5,2)  node[]{$Q_a$}
(2,0) node[]{$Q_b$}
;
\end{circuitikz}
	\caption{Feedback interconnection of PNP and NPN 
		transistors showing differential
	negative resistance between terminals $P_{1}$ and $P_{0}$.}
	\label{fig:circuit:nr}
\end{figure}
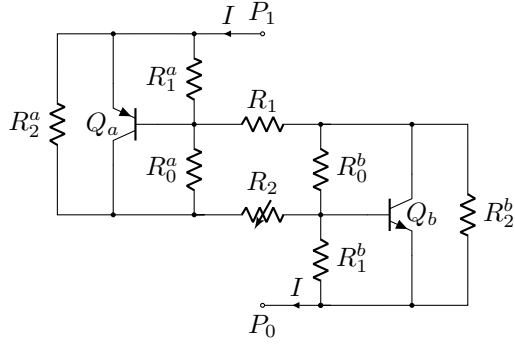

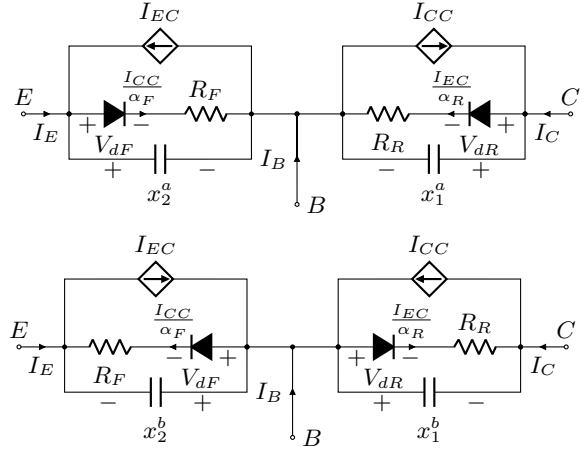
\begin{figure}[t]
	\centering
	\begin{circuitikz}[/tikz/circuitikz/bipoles/length=0.65cm, scale=0.6, american]
\tikzset{font=\small}
 \draw[color=black]



(0, 0) node[above]{$C$} to[short, o-*, i=$I_C$] ++(-1,0)
to[D*, i_=$\frac{I_{EC}}{\alpha_{R}}$, v^=$V_{dR}$] ++(-2,0) to[R, R=$R_R$]++(-2,0)
++(4,0) to[short, *-]++(0,-1) to[C, v^=$x_1^a$]++(-4,0) to[short, -*] ++(0,1)
-- ++(0,1.5) to[cI, l=$I_{CC}$] ++(4,0) -- ++(0,-1.5)
++(-4,0) to[short, *-*] ++(-1,0)
to[short, *-] ++(0, -1)
++(0,-1) node[right]{$B$} to[short, o-, i=$I_B$]++(0,2) to[short, -*]++(-1,0)
 -- ++(0,1.5) to[cI, l_=$I_{EC}$] ++(-4,0) -- ++(0,-1.5)
to[D*, i=$\frac{I_{CC}}{\alpha_F}$, v=$V_{dF}$] ++(2,0) to[R, R=$R_F$]++(2,0) 
++(-4,0) to[short, *-]++(0,-1) to[C, v=$x_2^a$] ++(4,0)
-- ++(0,1)
++(-5,0) node[above]{$E$} to[short, o-*, i_=$I_{E}$] ++(1,0)

;
\end{circuitikz}
	\begin{circuitikz}[/tikz/circuitikz/bipoles/length=0.65cm, scale=0.6, american]
\tikzset{font=\small}
 \draw[color=black]



(0,0) node[above]{$C$} to[short, o-*, i=$I_C$]++(-1,0)
-- ++(0,1.5) to[cI, l_=$I_{CC}$]++(-4,0)--++(0,-1.5)
to[D*, i=$\frac{I_{EC}}{\alpha_R}$, v=$V_{dR}$]++(2,0) to[R, R=$R_R$]++(2,0)
++(-4,0) to[short, *-] ++(0,-1) to[C, v=$x_1^b$]++(4,0) to[short, -*]++(0,1)
++(-4,0) to[short, *-*]++(-1.0,0)
++(0,-2) node[right]{$B$} to[short, i=$I_B$, o-]++(0,2)
to[short, -*]++(-1.0,0) 
to[D*, i_=$\frac{I_{CC}}{\alpha_{F}}$, v^=$V_{dF}$]++(-2,0) to[R, R=$R_F$]++(-2,0)
++(4,0) to[short, *-]++(0,-1) to[C, v^=$x_2^b$] ++(-4,0)
-- ++(0,2.5) to[cI, l=$I_{EC}$] ++(4,0) -- ++(0,-1.5)
++(-5,0) node[above]{$E$} to[short, i_=$I_E$, o-*] ++(1,0)
;
\end{circuitikz}
	\caption{Ebers-Moll model of bipolar junction transistors. 
		PNP transistor (above), and NPN	transistor (below).}
	\label{fig:ebersMoll}
\end{figure}

Motivated by the work of Chua \emph{et al.} on negative resistance devices~\cite{chua1983},
we consider the circuit in Fig.~\ref{fig:circuit:nr}. The circuit topology is fairly general
as, by opening or short-circuiting some of the resistors, it is possible to recover many of the
circuits considered in~\cite{chua1983}. 
This circuit is chosen as a practically relevant
running example to illustrate the theory developed throughout the paper but all the results are general and many other systems,
including mechanical or biological ones, could have been chosen to this purpose.
The circuit contains two bipolar junction transistors (BJTs) that we describe using
a dynamical version of the Ebers-Moll model depicted in Fig.~\ref{fig:ebersMoll} 
(see, e.g., \cite{getreu1978}), where each capacitor models the charge 
storage effects of the P-N and N-P junctions. 
Each diode in Fig.~\ref{fig:ebersMoll} satisfies the complementarity condition
\begin{displaymath}
	0 \leq I_{d} \perp (V^{*}-V_{d}) \geq 0 \;,
\end{displaymath}
where $I_{d}$ is the current flowing through the diode, $V_{d}$ is the voltage across 
its terminals, and $V^{*}$ is the forward voltage of the diode (typically between $0.3$ V--$0.7$ V).
After the application of Kirchhoff's laws, we realize that the circuit
evolves according to~\eqref{eq:lcs:1}-\eqref{eq:lcs:3} with state $\xi = [x_{1}^{a}, x_{2}^{a}, 
x_{1}^{b}, x_{2}^{b}]^{\top} \in \RE^{4}$. Here, $x_{k}^{a}$ and $x_{k}^{b}$, 
$k \in \{1, 2\}$, are the voltages across the capacitors of the transistors 
$Q_{a}$ and $Q_{b}$, respectively (see Fig.~\ref{fig:ebersMoll}). The 
complementary variables are given by
$z = [I_{EC}^{a}, I_{CC}^{a}, I_{EC}^{b}, I_{CC}^{b}] \in \RE^{4}$
and $w = [V^{*} - V_{dR}^{a}, V^{*}-V_{dF}^{a}, V^{*} - V_{dR}^{b}, V^{*} - V_{dF}^{b}]^{\top} \in \RE^{4}$.
We have the parameters $s = V^{*}$ and $r = P_{1} - P_{0}$, the latter being the potential difference
across the terminals shown in Fig.~\ref{fig:circuit:nr}. The remaining parameters
are
\begin{equation} \label{eq:lcs:param:1}
\begin{aligned}
 A &= Q
  \begin{bmatrix}
  	A_{1,1} & A_{1,2}
  	\\
	A_{1, 2}^{\top} & A_{2,2}
 \end{bmatrix}, \quad B =  Q
 \begin{bmatrix}
	T & 0
	\\
	0 & T
 \end{bmatrix}, \\
 E_{1} &=  Q [-G_{2}, G_{1} + G_{2}, -G_{1}, G_{1} + G_{2}]^{\top}, \quad E_{2} = \mathbf{1}_{4}, \\
 Q &= \diag\left( \frac{1}{C_{1}^{a}}, \frac{1}{C_{2}^{a}}, \frac{1}{C_{1}^{b}}, 
   \frac{1}{C_{2}^{b}}\right), \quad C  = -I_{4}, \\
 D &= \diag \left( \frac{R_{R}}{\alpha_{R}}, \frac{R_{F}}{\alpha_{F}}, 
   \frac{R_{R}}{\alpha_{R}}, \frac{R_{F}}{\alpha_{F}} \right) ,
 \end{aligned}
\end{equation}
where $\mathbf{1}_{4} \in \RE^4$ is a vector of ones and
\begin{equation} \label{eq:lcs:param:2}
\begin{aligned}
	A_{1,1} & = 
	\begin{bmatrix}
		- (G_{0}^{a} + G_{2}^{a} + G_{2}) 
		& G_{2}^{a} + G_{2}
		\\
		G_{2}^{a} + G_{2} 
		& -(G_{1}^{a} + G_{2}^{a} + G_{1} + G_{2})
	\end{bmatrix}, \\
	A_{2, 2} & = 
	\begin{bmatrix}
		-(G_{0}^{b} + G_{2}^{b} + G_{1}) 
		& G_{2}^{b} + G_{1}
		\\
		G_{2}^{b} + G_{1} 
		& -(G_{1}^{b} + G_{2}^{b} + G_{1} + G_{2})
	\end{bmatrix}, \\
	A_{1,2} & = 
	\begin{bmatrix}
		0 & G_{2}
		\\
		G_{1} & - (G_{1} + G_{2})
	\end{bmatrix}, \quad 
	T = 
	\begin{bmatrix}
		-\frac{1}{\alpha_{R}} & 1
		\\
		1 & -\frac{1}{\alpha_{F}}
	\end{bmatrix},
\end{aligned}
\end{equation}
with $G_{k}^{j} = 1/R_{k}^{j}$, $k \in \{1,2\}$, $j \in \{a, b\}$. Finally, 
$\alpha_{F}$, $\alpha_{R} \in (0,1)$ are the forward and reverse gains of the bipolar transistors.
\section{Bifurcations in linear complementarity systems} \label{sec:bifurcations}
In this section, we recall some facts about the solutions to LCPs and propose a procedure 
for detecting bifurcations in LCSs.
\subsection{Background on LCPs}
Linear complementarity problems have a rich geometric structure that determines the 
properties of their solution set. In what follows, we introduce the necessary elements
for the geometric analysis of LCPs.
 
\begin{defn}
	\label{defn:complementary:cone}
	Given an $\lcp(M,q)$ and a subset $\alpha \subseteq \Zint{n}$. 
   The \emph{complementary matrix} $C_{M}(\alpha) \in \RE^{n \times n}$ is the matrix
   whose columns are
	\begin{equation}
		\label{eq:complementary:matrix}
		C_{M}(\alpha)_{\bscdot, j}
		=
		\begin{cases}
			-M_{\bscdot, j} & j \in \alpha
			\\
			I_{\bscdot, j} & j \notin \alpha
		\end{cases} \;.
	\end{equation}
	The cone generated by $C_{M}(\alpha)$, $\pos C_{M}(\alpha)$, is called a 
	\emph{complementary cone} of $M$.
\end{defn}
 
A \emph{facet} of $\pos C_M(\alpha)$ is an $(n-1)$-dimensional face of the form
$$\pos C_M(\alpha)_{\bscdot, i^c}$$ for some $i \in \Zint{n}$ and $i^c$ the complement
of $\{i\}$ in $\Zint{n}$.
In what follows,  $\mathcal{K}(M)$ denotes the union of all facets of the complementary cones
of $M \in \RE^{n \times n}$.
The complementary cone $\pos C_{M}(\alpha)$ is called \emph{non-degenerate} if
it is solid, that is, if $C_{M}(\alpha)$ is nonsingular. 
Otherwise, $\pos C_{M}(\alpha)$ is called \emph{degenerate}. 
\begin{defn}
 A matrix $M \in \RE^{n\times n}$ is said to be \emph{non-degenerate} if all 
 its complementary cones, $\pos C_M(\alpha)$, $\alpha \subseteq \Zint{n}$ 
 are non-degenerate.
\end{defn}
\begin{defn}[\cite{danao1994,doverspike1982}]
	\label{defn:r0}
   We say that $M \in \RE^{n \times n}$ \emph{belongs to the class $R_{0}$} if,
   for each $\alpha \subseteq \Zint{n}$, we have $\ker(C_{M}(\alpha)) \cap 
   \RE_{+}^{n} = \{0\}$.
\end{defn}
The previous definition implies that all the complementary cones are
strictly pointed, that is, that $M$ does not have zero columns nor complementary cones that
are subspaces~\cite[Thm. 6.1.19]{cottle}. 
Now we define the piecewise linear map $f_M: \RE^{n} \to \RE^{n}$ as
\begin{equation}
	\label{eq:f:piecewise:linear}
	f_{M}(x) := C_{-M}(\alpha) x \quad \text{for $x \in \pos C_{I}(\alpha)$} \;.
\end{equation}
Note that the cones $\pos C_I(\alpha)$ are the $2^n$ orthants in $\RE^n$, so $f_M$ is defined
orthant-wise and $\alpha \subseteq \Zint{n}$ is simply a placeholder used for indexing each orthant. Also, 
mark that 
\begin{equation} \label{eq:f:cones}
 f_M(\pos C_I(\alpha)) = \pos C_M(\alpha) \;.
\end{equation}
\begin{thm}[\cite{cottle}]
	\label{thm:lcp:z:x:equivalent}
	Let $z \in \RE_{+}^{n}$ be a solution to the $\lcp(M,q)$. 
	Then, $x = (M - I_{n})z + q$ is a solution to the equation
	\begin{equation}
		\label{eq:lcp:f}
		f_{M}(x) - q = 0 \;.
	\end{equation}
	Conversely, if $x \in \RE^{n}$ is a solution to \eqref{eq:lcp:f}, then 
	$z = [-x]^{+}$ is a solution to the $\lcp(M,q)$.
\end{thm}
It follows that the $\lcp(M,q)$ and the problem~\eqref{eq:lcp:f} are equivalent in the 
sense that we only need to know the solutions to one of them in order to know
the solutions to the other. 
The functional representation~\eqref{eq:lcp:f} lends itself to the application of degree theory.
\begin{defn}
	Let $M \in \RE^{n \times n}$ be an $R_{0}$-matrix and $x \in \RE^{n}$ be in 
	the interior of $\pos C_{I}(\alpha)$ for some $\alpha \subseteq \Zint{n}$.
	The \emph{index} of $f_{M}$ at $x$, denoted as $\ind_{M}(x)$, is given by
	\begin{displaymath}
		\ind_{M}(x) = \sgn (\det M_{\alpha, \alpha}) \;.
	\end{displaymath}
	By convention, if $\alpha = \emptyset$, then $\ind_{M}(x) = 1$.
\end{defn}
\begin{defn}
	Let $M \in \RE^{n \times n}$ be an $R_{0}$-matrix and let $q \in \RE^{n}$ be such
	that $f_{M}^{-1}(q)$ consists of finitely many points and $\ind_{M}(x)$ is 
	well-defined for all $x \in f_{M}^{-1}(q)$. The \emph{degree} of $f_{M}$ at $q$, 
	denoted as $\deg_{M}(q)$ is
	\begin{displaymath}
		\deg_{M}(q) = \sum_{x \in f_{M}^{-1}(q)} \ind_{M}(x) \;.
	\end{displaymath}
\end{defn}
Note that, for a given $\alpha \subseteq \Zint{n}$, the index of $f_{M}$ at $x$ is the same 
for all $x \in \interior \pos C_{I}(\alpha)$ so, there are as many indices as there are 
subsets in $\Zint{n}$. For matrices in the class $R_{0}$, the degree is the same for any 
$q \notin \mathcal{K}(M)$~\cite[Theorem 6.1.14]{cottle}. Thus, for $R_{0}$-matrices the degree 
is global and one only needs to test one point $q \notin \mathcal{K}(M)$. The reader is addressed
to~\cite{cottle,garcia1981,howe1981} and references therein for a detailed 
account of degree theory in the context of complementarity problems.
The solutions to the $\lcp(M,q)$ depend on the geometry of the complementary cones 
$\pos C_{M}(\alpha)$. More precisely, there exists at least one solution $x$ 
of~\eqref{eq:f:piecewise:linear} for every $\alpha \subseteq \Zint{n}$ such
that $q \in \pos C_{M}(\alpha)$. If $C_{M}(\alpha)$ is nonsingular the associated solution 
is unique, whereas there exists a continuum of solutions if $C_{M}(\alpha)$ is singular. Thus, 
for a given $q$, there can be no solutions, there can be one solution, multiple isolated solutions, 
or a continuum of solutions, depending on how many complementary cones $q$ belongs to 
and what the properties of these cones are.
We will later make use of an alternative representation of $f_{M}$.
\begin{lem} \label{lemma:pwl:f:projection}
	The map~\eqref{eq:f:piecewise:linear} can also be written as
	\begin{equation}
		\label{eq:pwl:f:projection}
		f_{M}(x) = [x]^{+} - M [-x]^{+} \;.
	\end{equation}
\end{lem}
\subsection{LCS bifurcations are determined by LCP singularities}
The main issues addressed in the literature about LCS are well-posedness and
asymptotic behavior of the resulting trajectories (see, 
e.g.,~\cite{brogliato2003,brogliato2020,camlibel2002,heemels2000,schaft1998}).
If $D$ is a $P$-matrix (i.e., all its principal minors are positive) then, the inverse $f^{-1}_D$
is well-defined~\cite[Theorem 3.3.7]{cottle}, so for every $\xi$ and $s$, 
there exists a unique $z$ satisfying \eqref{eq:lcs:2}-\eqref{eq:lcs:3} and given by
\begin{equation} \label{eq:z}
	z = \left[ -f_D^{-1}(C\xi + E_2 s) \right]^+ \;.
\end{equation}
Substituting~\eqref{eq:z} in~\eqref{eq:lcs:1},~\eqref{eq:Sigma} can be written explicitly as
\begin{equation} \label{eq:explicit}
	\dot{\xi}(t) = F(\xi(t),r,s)
\end{equation}
with
\begin{equation}\label{eq:F def}
	F(\xi,r,s) = A \xi + B \left[ -f_D^{-1}(C\xi + E_2 s) \right]^+ + E_{1} r \;.
\end{equation}
Furthermore, it can be readily shown that $F$ is globally Lipschitz-continuous in $\xi$~\cite{gowda1992},
so that~\eqref{eq:Sigma} is well posed, i.e., its solutions
are global and unique. A less-explored problem is 
understanding how the equilibria of~\eqref{eq:Sigma} change, appear, and 
disappear as a function of the input parameters $r$ and $s$, that is, what is the nonsmooth equilibrium bifurcation 
structure of~\eqref{eq:Sigma}.
The following definition formalizes the notion of non-smooth equilibrium bifurcation. 
It uses a notion of regularity that is closely related to that proposed in 
\cite{robinson1980} and it is inspired by the notion of smooth equilibrium bifurcation 
used in~\cite{golubitsky}.
\begin{defn}
 Given a Lipschitz-continuous map depending on $m$ parameters, $g:\RE^n \times \RE^m\to \RE^n$, 
 a point $(y_0,p_0)$ is a \emph{regular solution} of $g$ if $g(y_0,p_0)=0$ and there exists a
 neighborhood $W=U \times V$ of $(y_0,p_0)$ such that the map $p \mapsto \{y \in U \mid g(y,p)=0 \}$ 
 is single-valued and Lipschitz-continuous over $V$\footnote{To simplify the exposition, a
 multi-valued map whose image is a singleton will be treated as a single-valued map.}. Solutions 
 of $g(y,p)=0$ that are not regular are called \emph{bifurcations} of $g$.
\end{defn}
Let $\partial_y g(y,p)$ denote the Clarke generalized Jacobian of $g$ with respect
to $y$ at $(y,p)$, that is,
\begin{displaymath}
	\partial_y g(y,p) = \co \Big\{ \lim_{i \to \infty} D_y g(y_i,p) \mid y_i \to y,
	y_i \not\in \Omega_g \Big\} \;,
\end{displaymath}
where $\Omega_g$ is the set of measure zero where the 
Jacobian $D_y g$ of the Lipschitz-continuous map $g$ does not exist. We say that $\partial_y g(y,p)$ 
is of \emph{maximal rank}, or \emph{nonsingular}, if every matrix in $\partial_y g(y,p)$ is nonsingular. 
Conversely, we say that $\partial_y g(y,p)$ is singular if it contains a singular matrix. The following
proposition is a direct consequence of the nonsmooth Implicit Function Theorem~\cite[p. 256]{clarke1990}.
\begin{prop}\label{prop: bifu to singu}
	If $(y_0,p_0)$ is a bifurcation of $g$ then $\partial_y g(y_0,p_0)$ is singular.
\end{prop}
The following theorem shows that equilibria of~\eqref{eq:Sigma} can be put in one-to-one 
correspondence with the solution of an associated LCP.
\begin{thm} \label{thm:equilibria}
 Consider~\eqref{eq:Sigma} with $D$ a $P$-matrix and $A$ nonsingular, and define
 \begin{subequations} \label{eq:lcp:Mq}
 \begin{align}
             M &= D - C A^{-1} B \label{eq:lcp:M} \\
  \bar{q}(r,s) &= E_{2}s - C A^{-1} E_{1} r \label{eq:lcp:q} \;.
 \end{align}
 \end{subequations}
 If $\xi^\ast$ is an equilibrium of~\eqref{eq:explicit}, then $x^\ast = X(\xi^\ast,s)$ with
 \begin{equation} \label{eq:xi to x}
  X(\xi,s) = f_D^{-1}(C\xi + E_2 s)
 \end{equation}
 is a solution of $f_M(x^\ast) = \bar{q}(r,s)$. Conversely, if $x^\ast$ is a solution of
 $f_M(x^\ast) = \bar{q}(r,s)$, then $\xi^\ast = \Xi(x^\ast,r)$ with
 \begin{equation} \label{eq:x to xi}
  \Xi(x,r)=-A^{-1}\left( B[-x]^++E_1 r \right)
 \end{equation}
 is an equilibrium of~\eqref{eq:explicit}.
\end{thm}
\begin{cor} \label{cor:bifs}
 Under the same assumptions and definitions used in Theorem~\ref{thm:equilibria}, 
 the point $(\xi_0,r_0,s_0)$ is a bifurcation of $F$ if, and only if,
 $(x_0,q_0)=(X(\xi_0,s_0),\bar{q}(r_0,s_0))$ is a bifurcation of 
 $(x,q) \mapsto f_M(x)-q$.
\end{cor}
It follows that the steady-state bifurcations of an LCS are characterized by the bifurcations of an associated LCP. Furthermore, by Proposition~\ref{prop: bifu to singu}, we can make two key observations. First, LCS bifurcations are non-hyperbolic equilibria of~\eqref{eq:Sigma}, in the sense that Jacobian $\partial_{\xi} F(\xi,r,s)$ at a bifurcation point contains matrices with some zero eigenvalues. As such, system~\eqref{eq:Sigma} can lose stability at bifurcations. Second, bifurcations of an LCP can be found constructively by looking for its {\em singularities}, i.e., points $(x_0,q_0)$ such that $f_M(x_0)-q_0=0$ and $\partial_{x} f_M(x_0)$ is singular. Observing that 
\begin{displaymath}
	\partial f_M(x)=\co\left\{ C_{-M}(\alpha)\mid x\in\pos C_I(\alpha) \right\}  \;,
\end{displaymath}
it follows that $\partial f_M(x)$ is a singleton whenever $x$ belongs to the interior 
of an orthant and a set of matrices whenever $x$ belongs to the common boundary of 
two or more orthants. 
Hence, for a point $(x_{0}, q_{0})$ to be a nonsmooth singularity of the $\lcp(M,q)$ it 
is necessary that, for some $\alpha \subseteq \Zint{n}$, either 
$x_{0} \in \boundary \pos C_{I}(\alpha)$ 
or $x_{0} \in \interior \pos C_{I}(\alpha)$ with $C_{-M}(\alpha)$ singular. 
In the former case $\partial f_{M}$ is a set for which the maximal rank condition may 
not hold, whereas in the latter case $\partial f_{M}$ is a singular matrix and therefore
it is not of maximal rank.
Note that, in both cases, $q_{0} \in \boundary C_{M}(\alpha)$.
Therefore, nonsmooth bifurcations in LCPs are essentially determined by the 
configuration of the complementarity cones and how $q$ moves across them. 
In practice, the \emph{path} traced by $q$ depends on control (or 
\emph{bifurcation}) parameters 
	$\lambda \in \RE^{l}$. That is, $q = \overline{q}(\lambda)$ where 
	$\overline{q}: \RE^{l} \to \RE^{n}$ is assumed continuous. Thereby, as $\lambda$ changes, 
the number of solutions of the LCP might change depending on the properties and the 
number of complementary cones that $\overline{q}(\lambda)$ traverses. A \emph{bifurcation diagram} is a
graphical representation of how equilibrium solutions change as a function of $\lambda$. 
The notion of equivalence for LCPs introduced in the next section helps in the task of analyzing
and classifying complementarity-cone configurations (and therefore LCP singularities and associated LCS bifurcations),
as illustrated through both applied and theoretical extended examples.
\section{Equivalence of LCPs} \label{sec:equivalence}
\begin{figure}
	\centering
	\includegraphics{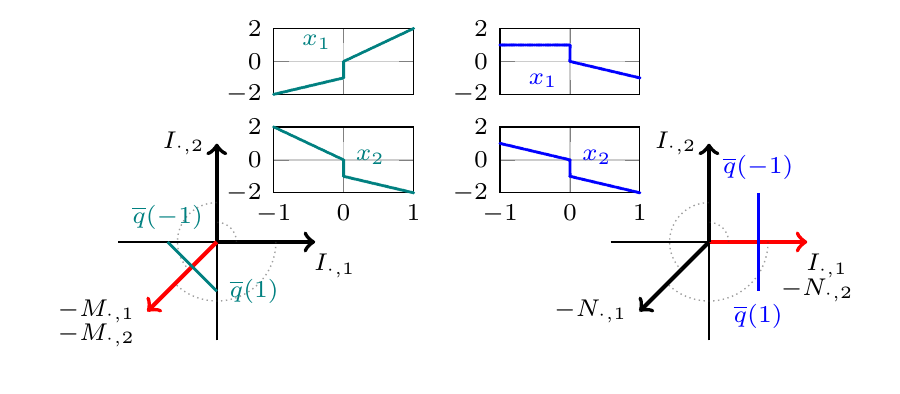}
	\caption{Complementary cones associated with the matrices $M$ (left) and $N$ (right) 
	from \eqref{eq:example:2:M:N}. Red rays denote degenerate complementary cones, 
whereas the line segments denote the path $\overline{q}$. The boxes show the
associated LCP bifurcation diagrams for the two affine paths indicated by the thin gray (left) and thin blue (right) line, respectively.}
	\label{fig:example:2:cones:M:N}
\end{figure}

Our notion of equivalence between an $\lcp(M,q)$ and an $\lcp(N,r)$ involves both topological
and algebraic properties. The algebraic properties deal with the relations among the 
complementary cones of $M$ and $N$. The relevant algebraic structure is that of a 
Boolean algebra, a subject that we now briefly recall (see~\cite{givant,sikorski} 
for more details).
\begin{defn}
	\label{defn:field:sets}
	Let $X$ be a set. A \emph{field of sets} is a pair $(X, \Fi)$, where 
   $\Fi \subseteq \mathcal{P}(X)$ is any non-empty family of sets that 
   is closed under the set operations of complement and finite union.
\end{defn}
It follows easily from the De Morgan rules that a field of sets $(X, \Fi)$ 
is also closed under finite intersections of sets. Note also that $\emptyset$ and $X$ 
are always members of $\Fi$. In the cases when the set $X$ is clear from the
context, we will denote the field of sets simply by the collection $\Fi$.
Fields of sets are concrete examples of Boolean algebras, see, e.g.,~\cite{sikorski}.
Therefore, the usual algebraic concepts apply to them.
\begin{defn}
	\label{defn:boolean:homeomorphism}
	Consider the fields $(X, \Fi)$ and $(\hat{X}, \hat{\Fi})$. 
   A mapping $h: \Fi \to \hat{\Fi}$ is said to be \emph{Boolean} 
   if, for all $P_{1}, P_{2} \in \Fi$,
	\begin{displaymath}
		h(P_{1} \bigcap P_{2}) = h(P_{1}) \bigcap h(P_{2})
	\end{displaymath}
	and
	\begin{displaymath}
		h(X \setminus P_{1}) = \hat{X} \setminus h(P_{1}) \;.
	\end{displaymath}
   A Boolean map that is also a bijection is said to be a \emph{Boolean isomorphism}. A
   Boolean isomorphism that maps a field of sets to itself is said to be a \emph{Boolean
   automorphism}.
\end{defn}
\begin{defn}
	\label{defn:boolean:map:induced}
   A map $\varphi_*: \Fi \to \hat{\Fi}$ is said to be
   \emph{induced} by a map $\varphi: \hat{X} \to X$ if, for every set 
   $P \in \Fi$,
	\begin{displaymath}
		\varphi_*(P) = \varphi^{-1}(P) \;.
	\end{displaymath}
\end{defn}
Let $\Ge \subseteq \mathcal{P}(X)$. The field of sets \emph{generated} by
$\Ge$, denoted $\Span \Ge$, is the intersection of all the fields of sets containing $\Ge$.
That is, it is the smallest field of sets containing $\Ge$.
In what follows, we consider the collection of complementary cones $\Ge_{M} = 
\{\pos C_{M}(\alpha)\}_{\alpha}$ and we denote the field of sets generated by 
$\Ge_{M}$ as $\Fi_{M}$.
We are now ready to introduce our main definition.
\begin{defn}
	\label{defn:lcp:equivalence}
	Two matrices $M, N \in \RE^{n \times n}$ are \emph{LCP-equivalent}, denoted
	$M \sim N$, if the fields of sets $\Fi_{M}$ and $\Fi_{N}$ are isomorphic
	and the Boolean isomorphism $\varphi_*: \Fi_{M} \to \Fi_{N}$ is induced
	by a homeomorphism $\varphi: \RE^{n} \to \RE^{n}$.
\end{defn}
In other words, two matrices are LCP-equivalent if the Boolean structures of their
associated complementary cones are isomorphic. That is, if there is a bijection
$\varphi_*: \Fi_{M} \to \Fi_{N}$ preserving intersections, unions, 
and complements of complementary cones. Since $\varphi_*$ is required to be induced by a 
homeomorphism, it suffices to verify that $\varphi$ induces a bijection from
$\Ge_{M}$ to $\Ge_{N}$~\cite[Cor. 12]{castanos2020}.
As an illustration, let us consider the matrices
\begin{equation}
	\label{eq:example:2:M:N}
	M = 
	\begin{bmatrix}
		1 & 1
		\\
		1 & 1
	\end{bmatrix} \quad \text{and} \quad N =
	\begin{bmatrix}
		1 & -1
		\\
		1 & 0
	\end{bmatrix} \;.
\end{equation}
The associated complementary cones and illustrative LCP bifurcation diagrams are depicted in Fig.~\ref{fig:example:2:cones:M:N}.
Note that, although $M$ is singular and $N$ is not, they are in fact LCP-equivalent.
To see this, consider the homeomorphism $\varphi: \RE^{2} \to \RE^{2}$,
\begin{displaymath}
	\varphi(q') = \begin{bmatrix}
		-1 & 0
		\\
		-1 & 1
	\end{bmatrix} q' \;.
\end{displaymath}
Simple but lengthy computations show that 
\begin{displaymath}
 \varphi_* : \pos C_{M}(\alpha) \mapsto \pos C_{N}(\beta(\alpha))
\end{displaymath}
with $\beta: \mathcal{P}(\Zint{n}) \to \mathcal{P}(\Zint{n})$ given by
\begin{displaymath}
	\emptyset \mapsto \{1\} \;,  \quad 
	   \{1\}  \mapsto \emptyset \;, \quad
	  \{2 \}  \mapsto \{1, 2\} \;, \quad
	\{1, 2\}  \mapsto \{2\} \;.
\end{displaymath}
That is, $\varphi$ induces a bijection $\Ge_{M} \to \Ge_{N}$ and hence
an isomorphism $\Fi_M \to \Fi_N$.
LCP-equivalence ensures that the two problems have similar bifurcation diagrams in all cases in which the bifurcation parameter paths in the two problems traverse isomorphic complementarity cones. This is the case of the example in Fig.~\ref{fig:example:2:cones:M:N}, which leads in both problems to a continuous bifurcation diagram characterized by the transition from one to an infinity (at the crossing of the degenerate cone) to one solution. The notion of equivalence between bifurcation diagrams will be formalized in future works.
\begin{thm} \label{thm:equivalence:suf}
	Consider two matrices $M, N \in \RE^{n \times n}$ and suppose there exists
 	homeomorphisms $\varphi, \psi: \RE^{n} \to \RE^{n}$ such that
	\begin{equation}
	\label{eq:equiv:f_M:f_N}
		f_{M} = \varphi \circ f_{N} \circ \psi \;,
	\end{equation}
	where $\psi$ induces a Boolean automorphism on $\Fi_{I}$. Then, $M \sim N$.
\end{thm}
There is a slightly weaker converse result.
\begin{thm} \label{thm:equivalence:nec}
	Consider two non-degenerate matrices $M, N \in \RE^{n \times n}$ and suppose 
   that $M \sim N$ with homomeorphism $\varphi$. Then, there exists another 
   homeomorphism $\psi: \RE^{n} \to \RE^{n}$ that induces an automorphism
   $\psi_*: \Fi_I \to \Fi_I$ and such that~\eqref{eq:equiv:f_M:f_N} holds.
\end{thm}
Condition~\eqref{eq:equiv:f_M:f_N} is the commutative diagram
\begin{displaymath}
\begin{tikzcd}
 \RE^n \arrow{r}{\psi} \arrow[swap]{d}{f_M} & \RE^n \arrow{d}{f_N} \\
 \RE^n                                      & \arrow{l}{\varphi} \RE^n
\end{tikzcd} \;.
\end{displaymath}
It is standard in the literature of singularity theory~\cite{arnold}
and ensures that we can continuously map solutions of the problem $f_M(x) = q$
into solutions of the problem $f_N(x') = \varphi^{-1}(q)$. The requirement on $\psi$ being a
Boolean automorphism implies that $\psi$ maps orthants into orthants, intersections of 
orthants into intersections of orthants, and so forth; and this 
ensures that the complementarity condition is not destroyed by the homeomorphisms.
Linear homeomorphisms preserving the complementarity conditions include permutations of
coordinates and dilations of the coordinate axes~\cite{howe1981}.
\begin{cor} \label{cor:equivalent:permutation}
	Let $P \in \RE^{n \times n}$ be a permutation matrix. Then, for any matrix $M \in \RE^{n\times n}$,
   $M \sim N = P^{\top} M P$.
\end{cor}
\begin{cor} \label{cor:equivalent:diagonal}
	Let $D \in \RE^{n \times n}$ be a diagonal matrix with strictly positive entries.
	Then, for any matrix $M \in \RE^{n\times n}$, $M \sim D^{-1} M D$ and 
	$M \sim M D$.
\end{cor}
\begin{defn}
	Let $M \in \RE^{n \times n}$ and $\beta \subseteq \Zint{n}$
	be such that $M_{\beta, \beta}$ is nonsingular. The \emph{principal
	pivotal transform} (PPT) of $M$ relative to $\beta$ is the matrix 
	$N \in \RE^{n\times n}$ defined by
	\begin{align*}
		N_{\beta, \beta} & = \left(M_{\beta, \beta}\right)^{-1}
		\\
		N_{\beta, \beta^{c}} & = -\left(M_{\beta, \beta}\right)^{-1} 
		M_{\beta, \beta^{c}}
		\\
		N_{\beta^{c}, \beta} & = M_{\beta^{c}, \beta} \left(M_{\beta, \beta}
		\right)^{-1}
		\\
		N_{\beta^{c}, \beta^{c}} & = M_{\beta^{c}, \beta^{c}} - M_{\beta^{c}, \beta}
		\left(M_{\beta, \beta}\right)^{-1} M_{\beta, \beta^{c}}
	\end{align*}
	By convention, the PPT of $M$ relative to $\emptyset$ is $M$.
	\label{defn:ppt:M}
\end{defn}
\begin{cor} \label{cor:ppt:equivalence}
	Let $N$ be the PPT of $M$ with respect to $\beta$. Then,	$M \sim N$.
\end{cor}
\section{Stability of LCPs} \label{sec:stability}
Given an $\lcp(M,q)$,  we are now concerned with characterizing stability of the structure of the
complementary cones with respect to changes in $M$ and, therefore, the stability of its nonsmooth equilibrium bifurcations.
\subsection{Definition and characterization of stability}
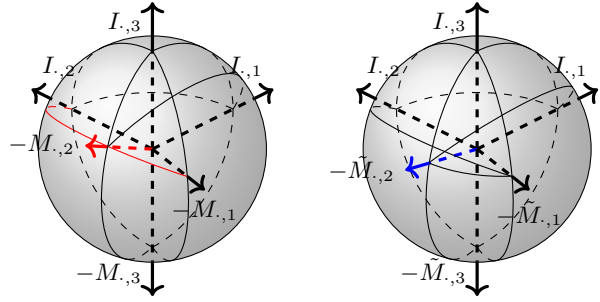
\begin{figure}
	\centering
 	\tdplotsetmaincoords{60}{-45}

\begin{tikzpicture}[scale=1.5, tdplot_main_coords]
\pgfmathsetmacro{\thF}{75.5}
\pgfmathsetmacro{\fhF}{-116.5650}
\pgfmathsetmacro{\thS}{61.5}
\pgfmathsetmacro{\fhS}{-161.5650}
\pgfmathsetmacro{\thT}{180}
\pgfmathsetmacro{\fhT}{-180}

\tdplotsetcoord{O}{0.0}{0.0}{0.0}

\tdplotsetcoord{M1}{1}{\thF}{\fhF}
\tdplotsetcoord{M2}{1}{\thS}{\fhS}
\tdplotsetcoord{M3}{1}{\thT}{\fhT}

\tdplotsetcoord{N1}{1.5}{\thF}{\fhF}
\tdplotsetcoord{N2}{1.5}{\thS}{\fhS}
\tdplotsetcoord{N3}{1.5}{\thT}{\fhT}

\shade[ball color = gray!30, opacity = 0.5] (0,0) circle (1.0cm);
\draw (0,0) circle (1cm);

\draw[dashed, very thick, opacity=1.0] (0,0,0) -- (1.27,0,0);
\draw[dashed, very thick, opacity=1.0] (0,0,0) -- (0,1.27,0);
\draw[dashed, very thick, opacity=1.0]  (0,0,0) -- (0,0,1);

\draw[very thick,->, opacity=1.0] (1.27,0,0) -- (1.5,0,0) node[anchor=south east]{\small $I_{\cdot, 1}$};
\draw[very thick,->, opacity=1.0] (0,1.27,0) -- (0,1.5,0) node[color=black, anchor=south west]{\small $I_{\cdot, 2}$};
\draw[very thick,->, opacity=1.0]  (0,0,1) -- (0,0,1.5) node[anchor=north east]{\small $I_{\cdot, 3}$};

\draw[very thick, dashed, opacity=1.0] 
(0,0,0) -- (M1); 
\draw[very thick, dashed, color=red, opacity=1.0] 
(0,0,0)--(M2); 
\draw[very thick, dashed, opacity=1.0] 
(0,0,0)--(0,0,-1.15); 

\draw[very thick,->, opacity=1.0] 
(M1) -- (N1) node[color=black, anchor=north]{\small $-M_{\cdot, 1}$};
\draw[very thick,->, color=red, opacity=1.0] 
(M2)--(N2) node[color=black, anchor=east]{\small $-M_{\cdot, 2}$};
\draw[very thick,->, opacity=1.0] 
(0,0,-1.15)--(N3) node[anchor=south east]{\small $-M_{\cdot, 3}$};

\tdplotsetthetaplanecoords{0}
\draw[tdplot_rotated_coords, thin, opacity=1.0] (1,0,0) arc (0:35:1);
\draw[tdplot_rotated_coords, thin, opacity=1.0, dashed] ({cos(35)},{sin(35)}, 0) arc (35:90:1);

\tdplotsetthetaplanecoords{90}
\draw[tdplot_rotated_coords, thin, opacity=1.0] (1,0,0) arc (0:35:1);
\draw[tdplot_rotated_coords, thin, opacity=1.0, dashed] ({cos(35)},{sin(35)}, 0) arc (35:90:1);

\tdplotsetrotatedcoords{0}{0}{0}
\draw[tdplot_rotated_coords, thin, opacity=1.0, dashed] (1,0,0) arc (0:90:1);

\tdplotsetrotatedcoords{0}{-180+30}{0}
\draw[tdplot_rotated_coords, thin, color=red, opacity=1.0, dashed] (0,1,0) arc (90:50:1);
\draw[tdplot_rotated_coords, thin, color=red, opacity=1.0] ({cos(50)},{sin(50)},0) arc (50:-60:1);

\tdplotsetthetaplanecoords{\fhF}
\draw[tdplot_rotated_coords, thin, opacity=1.0] (1,0,0) arc (0:\thF:1);

\tdplotsetrotatedcoords{-90}{-59.5}{0}
\draw[tdplot_rotated_coords, thin, opacity=1.0, dashed] (0,1,0) arc (90:50:1);
\draw[tdplot_rotated_coords, thin, opacity=1.0] ({cos(50)},{sin(50)},0) arc (50:-56:1);

\tdplotsetthetaplanecoords{\fhS}
\draw[tdplot_rotated_coords, thin, opacity=1.0] (1,0,0) arc (0:\thS:1);



\tdplotsetthetaplanecoords{90}
\draw[tdplot_rotated_coords, thin, opacity=1.0, dashed] (0,1,0) arc (90:180:1);
\tdplotsetthetaplanecoords{0}
\draw[tdplot_rotated_coords, thin, opacity=1.0, dashed] (0,1,0) arc (90:180:1);

\tdplotsetthetaplanecoords{\fhF}
\draw[tdplot_rotated_coords, thin, opacity=1.0, dashed] (-1,0,0) arc (180:150:1);
\draw[tdplot_rotated_coords, thin, opacity=1.0] ({cos(150)},{sin(150)},0) arc (150:75:1);

\tdplotsetthetaplanecoords{\fhS}
\draw[tdplot_rotated_coords, thin, opacity=1.0] (M2) arc (\thS:150:1);
\draw[tdplot_rotated_coords, thin, opacity=1.0, dashed] ({cos(150)},{sin(150)},0) arc (150:180:1);





\end{tikzpicture} 
 	\tdplotsetmaincoords{60}{-45}
\begin{tikzpicture}[scale=1.5, tdplot_main_coords]

\pgfmathsetmacro{\thF}{75.5}
\pgfmathsetmacro{\fhF}{-116.5650}
\pgfmathsetmacro{\thS}{70}
\pgfmathsetmacro{\fhS}{-161.5650}
\pgfmathsetmacro{\thT}{180}
\pgfmathsetmacro{\fhT}{-180}
\tdplotsetcoord{O}{0.0}{0.0}{0.0}

\tdplotsetcoord{M1}{1}{\thF}{\fhF}
\tdplotsetcoord{M2}{1}{\thS}{\fhS}
\tdplotsetcoord{M3}{1}{\thT}{\fhT}

\tdplotsetcoord{N1}{1.5}{\thF}{\fhF}
\tdplotsetcoord{N2}{1.5}{\thS}{\fhS}
\tdplotsetcoord{N3}{1.5}{\thT}{\fhT}

\shade[ball color = gray!30, opacity = 0.5] (0,0) circle (1.0cm);
\draw (0,0) circle (1cm);

\draw[dashed, very thick, opacity=1.0] (0,0,0) -- (1.27,0,0);
\draw[dashed, very thick, opacity=1.0] (0,0,0) -- (0,1.27,0);
\draw[dashed, very thick, opacity=1.0]  (0,0,0) -- (0,0,1);

\draw[very thick,->, opacity=1.0] (1.27,0,0) -- (1.5,0,0) node[anchor=south east]{\small $I_{\cdot, 1}$};
\draw[very thick,->, opacity=1.0] (0,1.27,0) -- (0,1.5,0) node[color=black, anchor=south west]{\small $I_{\cdot, 2}$};
\draw[very thick,->, opacity=1.0]  (0,0,1) -- (0,0,1.5) node[anchor=north east]{\small $I_{\cdot, 3}$};

\draw[very thick, dashed, opacity=1.0] 
(0,0,0) -- (M1); 
\draw[very thick, dashed, color=blue, opacity=1.0] 
(0,0,0)--(M2); 
\draw[very thick, dashed, opacity=1.0] 
(0,0,0)--(0,0,-1.15);

\tdplotsetthetaplanecoords{0}
\draw[tdplot_rotated_coords, thin, opacity=1.0] (1,0,0) arc (0:35:1);
\draw[tdplot_rotated_coords, thin, opacity=1.0, dashed] ({cos(35)},{sin(35)}, 0) arc (35:90:1);

\tdplotsetthetaplanecoords{90}
\draw[tdplot_rotated_coords, thin, opacity=1.0] (1,0,0) arc (0:35:1);
\draw[tdplot_rotated_coords, thin, opacity=1.0, dashed] ({cos(35)},{sin(35)}, 0) arc (35:90:1);

\tdplotsetrotatedcoords{0}{0}{0}
\draw[tdplot_rotated_coords, thin, opacity=1.0, dashed] (1,0,0) arc (0:90:1);

\tdplotsetrotatedcoords{0}{-180+30}{0}
\draw[tdplot_rotated_coords, thin, opacity=1.0, dashed] (0,1,0) arc (90:50:1);
\draw[tdplot_rotated_coords, thin, opacity=1.0] ({cos(50)},{sin(50)},0) arc (50:-60:1);

\tdplotsetthetaplanecoords{\fhF}
\draw[tdplot_rotated_coords, thin, opacity=1.0] (1,0,0) arc (0:\thF:1);

\tdplotsetrotatedcoords{-90}{-49.5}{0}
\draw[tdplot_rotated_coords, thin, opacity=1.0, dashed] (0,1,0) arc (90:50:1);
\draw[tdplot_rotated_coords, thin, opacity=1.0] ({cos(50)},{sin(50)},0) arc (50:-62:1);

\tdplotsetthetaplanecoords{\fhS}
\draw[tdplot_rotated_coords, thin, opacity=1.0] (1,0,0) arc (0:\thS:1);



\tdplotsetthetaplanecoords{90}
\draw[tdplot_rotated_coords, thin, opacity=1.0, dashed] (0,1,0) arc (90:180:1);
\tdplotsetthetaplanecoords{0}
\draw[tdplot_rotated_coords, thin, opacity=1.0, dashed] (0,1,0) arc (90:180:1);

\tdplotsetthetaplanecoords{\fhF}
\draw[tdplot_rotated_coords, thin, opacity=1.0, dashed] (-1,0,0) arc (180:150:1);
\draw[tdplot_rotated_coords, thin, opacity=1.0] ({cos(150)},{sin(150)},0) arc (150:75:1);

\tdplotsetthetaplanecoords{\fhS}
\draw[tdplot_rotated_coords, thin, opacity=1.0] (M2) arc (\thS:150:1);
\draw[tdplot_rotated_coords, thin, opacity=1.0, dashed] ({cos(150)},{sin(150)},0) arc (150:180:1);

\tdplotsetrotatedcoords{-90}{0.0}{-105}
\draw[tdplot_rotated_coords, thin, opacity=1.0] (M2) arc (20:65:1);

\draw[very thick,->, opacity=1.0] 
(M1) -- (N1) node[color=black, anchor=north]{\small $-\tilde{M}_{\cdot, 1}$};
\draw[very thick,->, color=blue, opacity=1.0] 
(M2)--(N2) node[color=black, anchor=east]{\small $-\tilde{M}_{\cdot, 2}$};
\draw[very thick,->, opacity=1.0] 
(0,0,-1.15)--(N3) node[anchor=south east]{\small $-\tilde{M}_{\cdot, 3}$};




\end{tikzpicture}
	\caption{Projection onto the unit ball of complementary cones of the matrix 
		$M$ defined in~\eqref{eq:example:partition:M}(left) and a perturbation 
		$\tilde{M}$ (right).
		The generator $-M_{\bscdot, 2} \in \pos [-M_{\bscdot, 1}, I_{\bscdot, 2}]$ 
	(in red). Thus, $M$ is weakly degenerate whereas $\tilde{M}$ is not.}
	\label{fig:partition}
\end{figure}

Henceforth, we consider the usual topology on the vector space $\RE^{n \times n}$.
\begin{defn}
	A matrix $M \in \RE^{n \times n}$ is \emph{LCP-stable} if there exists a neighborhood
   $U$ of $M$ such that $M \sim N$ for all $N \in U$.
	\label{defn:lcp:stable}
\end{defn}
In order to characterize LCP stability, we take a deeper look into the 
geometric properties of the facets  of the complementary cones.
Recalling that $\mathcal{K}(M)$ denotes the union of all facets of the complementary cones
of $M \in \RE^{n \times n}$. Then, $\mathcal{K}(M) = \bigcup_{\alpha} \boundary
C_{M}(\alpha)$ \cite[p. 511]{cottle} and it breaks down $\RE^{n}$ into 
a finite number of connected regions. 
\begin{defn}
	\label{defn:partition}
	Let $M \in \RE^{n \times n}$.  
	The \emph{partition induced by $M$}, denoted as $\Pa_{M}$, 	
	is the finite family of connected components of $\RE^{n} \setminus \mathcal{K}(M)$.
\end{defn}
A set $Q_{M}^{i} \in \Pa_{M}$ is called a \emph{cell} of $\Pa_{M}$. 
The number of cells, $r$, in the partition induced by $M$ depends on the number of solid 
complementary cones and how they intersect.
The following properties are easy to verify,
\begin{enumerate}[i)]
    \item $Q_{M}^{i} \bigcap Q_{M}^{j} = \emptyset$ for 
       $i \neq j$, $i,j \in \Zint{r}$,
 
    \item $\RE^{n} = \closure \bigcup_{i} Q_{M}^{i}$, \label{it:cover}
    \item $\mathcal{K}(M) = \bigcup_{i} \boundary Q_{M}^{i}$. \label{it:boundary}
\end{enumerate}
The cells of the partition are (non-necessarily convex) open cones 
delimited by the facets of the solid complementary cones of $M$. 
For $P$-matrices, the cells of the partition precisely agree with the
interior of the complementary cones of $M$, see e.g.,~\cite{samelson1958}.
\begin{lem} \label{lemma:complementary:partition:M}
	Let $M \in \RE^{n\times n}$ be an $R_{0}$ matrix and let $r$ be the number 
	of sets in $\Pa_{M}$.
   For every $\alpha \subseteq \Zint{n}$ there exists $J \subset \Zint{r}$ 
	such that either
	\begin{displaymath}
		\pos C_{M}(\alpha) = \closure \bigcup_{j \in J} Q_{M}^{j} 
	\end{displaymath}
   (when $\pos C_{M}(\alpha)$ has non-empty interior) or
   \begin{displaymath}
      \pos C_{M}(\alpha) \subset \bigcup_{j \in J} \boundary Q_{M}^{j}
   \end{displaymath}
   (when the interior of $\pos C_{M}(\alpha)$ is empty).
\end{lem}
\begin{cor} \label{cor:partition:cones} 
   $M \sim N$ if, and only if, there exists a bijection 
   $\hat{\varphi}_*:\Pa_M \to \Pa_N$ induced by a homeomorphism $\hat{\varphi}:\RE^{n} \to \RE^{n}$.
\end{cor}
Corollary~\ref{cor:partition:cones} implies that, if $M \sim N$, then $\Pa_{M}$
and $\Pa_{N}$ must have the same number of cells.
Consider a matrix $M \in \RE^{n\times n}$ and a vector $-M_{\bscdot, k}$ with $k \in \Zint{n}$,
and let $\mathcal{T}_k(M)$ be the collection of facets for 
which $-M_{\bscdot, k}$ is not a generator,
\opt{twoside}{
 \begin{multline*}
  \mathcal{T}_{k}(M) := \Big\{ S \subset \RE^{n} \mid S = \pos C_{M}(\alpha)_{\bscdot, i^{c}}, \\
 	\alpha \subseteq \Zint{n}, \; k \notin \alpha, \; i \in \Zint{n}\Big\} \;.
 \end{multline*}
}
\opt{onecolumn}{
 \begin{displaymath}
  \mathcal{T}_{k}(M) := \Big\{ S \subset \RE^{n} \mid S = \pos C_{M}(\alpha)_{\bscdot, i^{c}}, \;
 	\alpha \subseteq \Zint{n}, \; k \notin \alpha, \; i \in \Zint{n}\Big\} \;.
 \end{displaymath}
}
\begin{defn}
 Given $M \in \RE^{n \times n}$, we say that $M$ is \emph{weakly degenerate} if
 it has a degenerate complementary cone or if 
 \begin{equation}
	 \label{eq:weak:degenerate:condition}
  -M_{\bscdot, k} \in \bigcup_{S \in \mathcal{T}_{k}(M)} S 
 \end{equation}
 for some $k \in \Zint{n}$.
\end{defn}
As an illustration, consider the matrix
\begin{equation}
	\label{eq:example:partition:M}
	M = 
	\begin{bmatrix}
		\frac{1}{2} & \frac{5}{3} & 0
		\\
		1 & 1 & 0
		\\
		-\frac{3}{10} & -1 & 1
	\end{bmatrix} \;.
\end{equation}
It is lengthy but straightforward to verify that there are no degenerate complementary cones,
so $M$ is non-degenerate. However, note that
\opt{twoside}{
 \begin{multline*}
 	\mathcal{T}_{2}(M) = \Big\{ 
 		\pos [I_{\bscdot, 2}, I_{\bscdot, 3}], 
 		\pos [I_{\bscdot, 2}, -M_{\bscdot, 3}],  
 		\pos [I_{\bscdot, 1}, I_{\bscdot, 3}], 
 		\\
 		\pos [-M_{\bscdot, 1}, I_{\bscdot, 3}], 
 		\pos [I_{\bscdot, 1}, -M_{\bscdot, 3}],
 		\\
 		\pos [-M_{\bscdot, 1}, -M_{\bscdot, 3}], 
 		\pos [I_{\bscdot, 1}, I_{\bscdot, 2}], 
  		\pos [-M_{\bscdot, 1}, I_{\bscdot, 2}]
 		\Big\}
 \end{multline*}
}
\opt{onecolumn}{
\begin{multline*}
	\mathcal{T}_{2}(M) = \Big\{ 
		\pos [I_{\bscdot, 2}, I_{\bscdot, 3}], 
		\pos [I_{\bscdot, 2}, -M_{\bscdot, 3}],  
		\pos [I_{\bscdot, 1}, I_{\bscdot, 3}], 
		\pos [-M_{\bscdot, 1}, I_{\bscdot, 3}], 
		\\
		\pos [I_{\bscdot, 1}, -M_{\bscdot, 3}],
		\pos [-M_{\bscdot, 1}, -M_{\bscdot, 3}], 
		\pos [I_{\bscdot, 1}, I_{\bscdot, 2}], 
		\pos [-M_{\bscdot, 1}, I_{\bscdot, 2}]
		\Big\}
\end{multline*}
}
and that $-M_{\bscdot, 2} \in \pos [-M_{\bscdot, 1}, I_{\bscdot, 2}]$, 
so $M$ is a weakly degenerate matrix.
Fig.~\ref{fig:partition} depicts the partitions that $M$ and a perturbation $\tilde{M}$
induce on the unit ball.
\begin{lem} \label{lem:nowhereDense}
	The set of weakly degenerate matrices is nowhere dense in $\RE^{n \times n}$.
\end{lem}
\begin{thm} \label{thm:lcp:stable}
	A matrix $M \in \RE^{n \times n}$ is LCP-stable if, and only if,
	$M$ is not weakly degenerate.
\end{thm}
\subsection{Explicit conditions for stability}
Weak degeneracy is connected to the \emph{positive} linear dependence property
of particular subsets of generators of $M$, which 
is difficult to verify for large matrices \cite{regis2016}. On the other hand, efficient
algorithms for verifying linear dependence are readily available. 
The fact that linear independence implies positive independence motivates the following.
\begin{cor} \label{cor:lcp:stable:principal:minor}
	If all minors of $M \in \RE^{n \times n}$ are different from zero, 
   then $M$ is LCP-stable.
\end{cor}
Note that the condition of the corollary is sufficient but not necessary. For example,
 the identity $I_n$ is stable, yet more than one of its minors are zero.
\subsection{Stability margin}
Let $\dist(\cdot, S): \RE^{n} \to \RE_{+}$ be the classical Euclidean 
distance from a point to a closed convex set $S \subset \RE^{n}$, and let $\lin A$ be 
the linear hull (span) of the columns of $A$.
\begin{defn}
	The stability margin of $M$, $\sm: \RE^{n \times n} \to \RE_{+}$, is given as,
	\begin{equation} \label{eq:stab:margin}
		\sm(M) = \min \left\{ d \mid d \in \mathcal{A} \cup \mathcal{B} \right\} \;,
	\end{equation}
	where
	\begin{align*}
		\mathcal{A} & = \left\{ \dist(-\overline{M}_{k}; S) \mid S \in \mathcal{T}_{k}(M), k \in \Zint{n} \right\} \;, \\
		\mathcal{B} & = \left\{ \dist(-\overline{M}_{k}; \lin C_{M}(\alpha)_{\cdot, k^{c}}) \mid \alpha \subseteq \Zint{n}, k \in \Zint{n} \right\} \;,
	\end{align*}
	and the column $\overline{M}_{\cdot, k} = \frac{1}{\Vert M_{\cdot, k} \Vert} M_{\cdot, k}$ is 
	taken as a point in $\RE^{n}$.
\end{defn}
Note that $\mathcal{A}$ contains zero elements whenever~\eqref{eq:weak:degenerate:condition} holds, whereas
$\mathcal{B}$ contains zero elements whenever there are degenerate complementary cones.
Thus, $\sm(M) = 0$ if, and only if, $M$ is not LCP-stable. It follows from the definition that, for any $M \in \RE^{n \times n}$, 
$\sm(M) \in [0, 1]$ and $\sm(M) =1$ if, and only if, $M = \diag\{a_{1}, \dots, a_{n}\}$, where $a_{i} > 0$ for $i \in \Zint{n}$.
\begin{exmp} \label{examp:sm}
	Consider the matrix
	\begin{equation}
		M(\varepsilon) = 
		\begin{bmatrix}
			-1 + \varepsilon & \varepsilon
			\\
			\varepsilon & -1+\varepsilon
		\end{bmatrix} \;.
		\label{eq:example:sm}
	\end{equation}
	Simple computations lead us to
	\begin{align*}
		\mathcal{A} & = \left\{ 1, \frac{\vert \varepsilon \vert}{\sqrt{(1 - \varepsilon)^{2} + \varepsilon^{2})}} \right\} \;, \\
		\mathcal{B} & = \left\{ \frac{\vert 1 - \varepsilon \vert}{\sqrt{(1 - 
			\varepsilon)^{2} + \varepsilon^{2}}}, \sin \left( \arccos\left( 
			\frac{2 \varepsilon(1 - \varepsilon)}{(1- \varepsilon)^{2} + 
         \varepsilon^{2}} \right) \right) \right\} \;,
	\end{align*}
	where $\arccos: [-1, 1] \to [0, \pi]$. Note that $\sm(M(\varepsilon)) = 0$ for 
	$\varepsilon \in \left\{0, 1/2, 1\right\}$ and also $\lim_{\varepsilon \to \pm \infty}\sm(M(\varepsilon)) = 0$ which indeed 
	characterize all unstable cases for $M(\varepsilon)$. Fig. \ref{fig:sm} below 
	depicts the stability margin of \eqref{eq:example:sm} as a function of $\varepsilon$. 
The stability margin~\eqref{eq:stab:margin} can also be used for defining the \emph{most stable matrices}. It can be
seen in Fig.~\ref{fig:sm} that, for the matrix in~\eqref{eq:example:sm}, local maxima are attained at 
$\varepsilon \in \{-1.37, 0.37, 0.64, 2.37\}$. 
It is shown in Section~\ref{sec:classification} below that such points indeed coincide with
the centers of mass of the stable regions presented in the classification diagram
of Fig.~\ref{fig:bifDiag}.
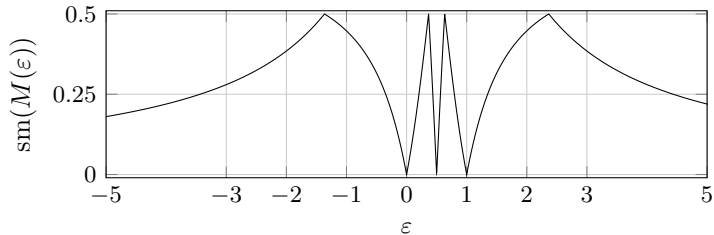
\begin{figure}[htpb]
		\centering
		\begin{tikzpicture}
	\pgfplotsset{height=0.3\textwidth, width=0.75\textwidth}

\begin{axis}
	[	xtick = {-5, -3, -2, -1, 0, 1, 2, 3, 5}, 
		xticklabels = {\small $-5$, \small $-3$, \small $-2$, \small $-1$, \small $0$, \small $1$, \small $2$, \small $3$, \small $5$}, 
	x grid style={white!80.0!black},
	xmajorgrids,
	xmin=-5, xmax=5,
	xlabel={\small $\varepsilon$},
	ytick = {0 , 0.25, 0.5},
	yticklabels = {\small $0$, \small $0.25$, \small $0.5$},
	y grid style={white!80.0!black},
	ylabel={$\textrm{sm}(M(\varepsilon))$},
	ymajorgrids,
	ymin=-0.01, ymax=0.51,
	ylabel style={at={(axis description cs:-0.1,.5)},anchor=south},
	xlabel near ticks,
]
\addplot[mark=none] table [x=X, y=Y, col sep=comma]{stabMargin3.csv};
\end{axis}
\end{tikzpicture}
		\caption{Stability margin of the matrix $M(\varepsilon)$ in 
			\eqref{eq:example:sm} with local minima at $\varepsilon \in 
			\{0, 1/2, 1\}$ (corresponding to unstable configurations),
			and local maxima at $\varepsilon \in \{-1.37, 0.37, 0.64, 2.37\}$.  }
		\label{fig:sm}
\end{figure}
\end{exmp}
\section{Application to negative-resistance circuits} \label{sec:application}
We will apply the results of the previous sections to the analysis of the circuit
described in Section~\ref{sec:circuit_model}. The application is motivated by~\cite{chua1983}, 
where the authors present a heuristic procedure for constructing negative differential-resistance 
devices using only BJTs and resistors. The design strategy outlined in~\cite{chua1983}
ensures the satisfaction of an analytic condition that is necessary for having multiple solutions.
Still, the question of whether a particular circuit actually exhibits negative resistance is 
settled by simulations. In this section, we employ the LCP formalism to derive necessary and 
sufficient conditions for having a negative resistance.
Our objective is to find values of parameters that 
guarantee bistability of the network, resulting in the hysteretic behavior intrinsic of differential
negative-resistance devices.
\subsection{Circuit equilibria}
Recall that the equilibria of~\eqref{eq:Sigma} are characterized by the solutions to the $\lcp(M, q)$,
where $M$ and $q$ are given in~\eqref{eq:lcp:M} and~\eqref{eq:lcp:q}. The matrix inverse
in~\eqref{eq:lcp:M} complicates the computation of equilibria and the ensuing bifurcation
analysis. Fortunately, the results of the previous sections ease the computational burden and open the
way for an analytical treatment.
Let $A, B, C \in \RE^{4\times 4}$ be nonsingular and defined by~\eqref{eq:lcs:param:1} 
and~\eqref{eq:lcs:param:2}, and let $N := -C A^{-1} B$. By applying a pivotal transformation to $N$ 
with $\beta = \{1, 2, 3, 4\}$ we obtain
\begin{displaymath}
 \hat{M} := -B^{-1} A C^{-1} \;.
\end{displaymath}
By Corollary~\ref{cor:ppt:equivalence}, $N \sim \hat{M}$. 
Let $M$ and $q$ be defined by~\eqref{eq:lcp:M},~\eqref{eq:lcp:q},~\eqref{eq:lcs:param:1} and~\eqref{eq:lcs:param:2}.
If $N$ is stable, $M = N + D \sim \hat{M}$ for $\|D\|$ small enough. Moreover, the solutions of the $\lcp(M,q)$ 
can be continuously mapped into the solutions of the $\lcp(\hat{M},\hat{q})$ with 
\begin{displaymath}
 \hat{q} := B^{-1} \left( A C^{-1} E_{2} s - E_{1} r \right) \;,
\end{displaymath}
where the expression follows from $\hat{q} = \varphi^{-1}(q)$ with $\varphi$ as in~\eqref{eq:varphi:pivot}.
The interest of $M \sim \hat{M}$ is that we no longer require to compute $M$, which involves
the inverse of $A$, defined by~\eqref{eq:lcs:param:1} and~\eqref{eq:lcs:param:2}. Instead, we require to compute
$\hat{M}$, which is easy, since $B$ is block diagonal and $C = -I_{4}$.  Indeed, $\hat{M}$ is given explicitly
by the submatrices
\begin{equation} \label{eq:lcp:circuit:M}
\begin{aligned}
	\hat{M}_{1,1} &=
    \begin{bmatrix}
		\frac{G_{0}^{a} + (1-\alpha_{F}) (G_{2}^{a} + G_{2})}{\alpha_{F}} & 
		\frac{\alpha_{F} (G_{1}^{a} + G_{1}) - (1 - \alpha_{F})(G_{2}^{a} + G_{2})}{\alpha_{F}} \\
		\frac{\alpha_{R} G_{0}^{a} - (1 - \alpha_{R})(G_{2}^{a} + G_{2})}{\alpha_{R}} &
		\frac{G_{1}^{a} + G_{1} + (1 - \alpha_{R})(G_{2}^{a} + G_{2})}{\alpha_{R}}
    \end{bmatrix} \;, \\
	\hat{M}_{1,2} &=
    \begin{bmatrix}
		-G_{1} &	\frac{\alpha_{F} G_{1} - (1 - \alpha_{F})G_{2}}{a_{F}} \\
		-\frac{G_{1}}{\alpha_{R}}	& \frac{G_{1} + (1 - \alpha_{R}) G_{2}}{\alpha_{R}}
    \end{bmatrix} \;, \quad
	\hat{M}_{2,1} =
    \begin{bmatrix}
		-G_{2} & \frac{\alpha_{F} G_{2}- (1 - \alpha_{F})G_{1}}{\alpha_{F}} \\
		-\frac{G_{2}}{\alpha_{R}} & \frac{G_{2} + (1 - \alpha_{R}) G_{1}}{\alpha_{R}}
    \end{bmatrix} \;, \\
	\hat{M}_{2,2} &=
    \begin{bmatrix}
		\frac{G_{0}^{b} + (1 - \alpha_{F}) (G_{2}^{b} + G_{1})}{\alpha_{F}} & 
      \frac{\alpha_{F} (G_{1}^{b} + G_{2}) - (1 - \alpha_{F})(G_{2}^{b} + G_{1})}{\alpha_{F}} \\
		\frac{G_{0}^{b} - (1 - \alpha_{R})(G_{2}^{b} + G_{1})}{\alpha_{R}} &
		\frac{G_{1}^{b} + G_{2} + (1 - \alpha_{R})(G_{2}^{b} + G_{1})}{\alpha_{R}}
    \end{bmatrix} \;. 
\end{aligned}
\end{equation}
\subsection{Circuit bifurcations}
We regard the potential difference $r = P_{1} - P_{0}$ as a bifurcation or control parameter. We wish to find a set of
resistor values for which, by changing $r$, the circuit transitions from global stability to bistability.
The following corollary gives sufficient conditions for attaining this objective.
\begin{cor}[{\cite[Corollary 4.3]{howe1981}}]
Let $\hat{M} \in \RE^{n \times n}$ be an $R_{0}$-matrix such that $\deg_{\hat{M}}(q) = 1$ for some
$q \in \RE^{n}$ and $f_{\hat{M}}$ has only one negative index. 
Then, there is $\alpha_{0} \subseteq \Zint{n}$ such that,
for any $\hat{q} \in \interior \pos C_{\hat{M}}(\alpha_{0})$, the
$\lcp(\hat{M},\hat{q})$ has exactly three solutions, whereas if $\hat{q} \in \RE^{n} 
\setminus \pos C_{\hat{M}}(\alpha_{0})$, the solution is unique.
\label{cor:hysteresis}
\end{cor}
With an explicit expression for $\hat{M}$, it is straightforward to check the conditions of Corollary~\ref{cor:hysteresis}.
 
\begin{prop} \label{prop:hysteresis}
 Let $\hat{M} \in \RE^{4\times 4}$ be given by~\eqref{eq:lcp:circuit:M} with all parameters finite and positive,
 and with $\alpha_F, \alpha_R \in (0,1)$. Define
 \begin{multline} \label{eq:gamma}
  \gamma := G_{0}^{a} G_{0}^{b} + (1 - \alpha_{F}) \left( G_{0}^{a} G_{2}^{b} + G_{0}^{b} G_{2}^{a} + (1 - \alpha_{F})G_{2}^{a} G_{2}^{b} \right) \\
	+ (1 - \alpha_{F}) \left( G_{0}^{a} + (1-\alpha_{F}) G_{2}^{a} \right) G_{1} \\
	- \left( (2\alpha_{F} - 1) G_{1} - (1-\alpha_{F}) \left( G_{0}^{b} + 
	(1-\alpha_{F}) G_{2}^{b} \right) \right) G_{2} \;.
 \end{multline}
 If:
 \begin{itemize}
  \item $\gamma > 0$, then the $\lcp(\hat{M},\hat{q})$ has a unique solution, regardless of $\hat{q}$.
  \item $\gamma < 0$, then the $\lcp(\hat{M},\hat{q})$ may have a unique solution 
	  or three solutions, depending
   on the specific value of $\hat{q}$.
  \item $\gamma = 0$, then $\hat{M}$ is LCP-unstable.
 \end{itemize}  
\end{prop}
For real transistors we have $\alpha_{F} \in [0.8, 0.99]$ and $\alpha_{R} \approx 0.5$~\cite{getreu1978}. 
It thus follows that we can make $\gamma < 0$ by setting $G_{1}$ and $G_{2}$ large enough, whenever 
the remaining conductances stay finite. Regarding the condition
$\hat{q} \in \interior \pos C_{\hat{M}}(\hat{\alpha}_{0})$ in Corollary~\ref{cor:hysteresis},
we will verify it numerically once the resistor parameters have been chosen. 
For concreteness, we consider all capacitors with the same value of $100\,\mu$F, 
$\alpha_{F} = 0.99$, $\alpha_{R} = 0.5$, and $s=0.7$ V. The values of the resistances $R_{k}^{j}$, 
$k \in \{0,1,2\}$, $j \in \{a, b\}$ are selected arbitrarily, whereas $G_{1}$ is chosen in such a way that
the factor associated with $G_{2}$ in~\eqref{eq:gamma} is strictly negative.
The chosen values are shown in Table~\ref{tab:param:value}.
 
\begin{figure}
	\centering
	\includegraphics{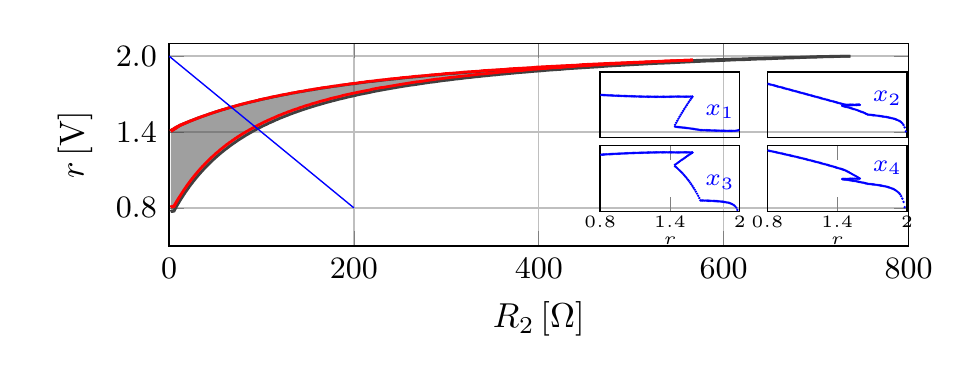}
	\caption{
		Two-parameter bifurcation diagram	
		in the $(R_{2},r)$-plane. 
      The open set $U_3$ where $\lcp(\hat{M}, \hat{q})$ has three solutions is highlighted in gray. 
      There is only one solution in $U_1= \interior \RE^2\setminus U_3$. The circuit
      bifurcates at $\boundary U_3$.
      The set of parameters $U_c$ for which the original $\lcp(M, q)$ bifurcates is shown in red.}
	\label{fig:lcp:region:feasible}
\end{figure}

\begin{table}[t]
	\centering
	\begin{tabular}{ccccc}
		$R_{0}^{a} = 100\,\Omega$ 
		& 
		$R_{1}^{a} = 2.2\, \mathrm{k}\Omega$ 
		& 
		$R_{2}^{a} = 100\, \Omega$
		\\
		\hline
		$R_{0}^{b} = 100\, \Omega$ 
		& 
		$R_{1}^{b} = 100\, \Omega$ 
		& 
		$R_{2}^{b} = 10\, \mathrm{k}\Omega$
		\\
		\hline
		& $R_{1} = 10\, \Omega$ & $R_{2} = 0\,\Omega - 1\, \mathrm{k} \Omega$
		\\
	\end{tabular}
	\caption{Values of the parameters used for the circuit of Fig.~\ref{fig:circuit:nr}.}
	\label{tab:param:value}
\end{table}
With the setting described in the previous paragraph, we can use $G_2$ to set the sign
of $\gamma$. Fig.~\ref{fig:lcp:region:feasible} shows {\color{blue} the $(R_{2},r)$-bifurcation diagrams
of $\lcp(\hat{M}, \hat{q})$ and $\lcp(M, q)$.} 
\subsection{Asymptotic properties of the circuit}
\begin{figure}
	\centering
	\begin{tikzpicture}
  \begin{groupplot}[group style={group size=1 by 3,vertical sep=4.5mm},
      height=3cm, width=0.75\textwidth, 
]
\nextgroupplot[xtick = {0, 2, 4, 6, 8, 10},
	xticklabels = {\small$0$, \small$2$,\small $4$, \small $6$, \small $8$, \small $10$},
x grid style={white!80.0!black},
xmajorgrids,
xmin=0.0, xmax=10,
ytick = {0.0, 0.7},
yticklabels = {\small $0$, \small $0.7$},
y grid style={white!80.0!black},
ylabel={$[\text{V}]$},
ymajorgrids,
ymin=-0.2, ymax=0.9,
]
\addplot [thick, black, forget plot] table [x=T, y=X1, col sep=comma]{circuit_simulationData.tex};
\addplot [thick, black, dashed, forget plot] table [x=T, y=X2, col sep=comma]{circuit_simulationData.tex};
\node[] at (axis cs: 0.5, 0.15, 0.16) {$\xi_{1}$};
\node[] at (axis cs: 0.5, 0.7, 0.16) {$\xi_{2}$};

\nextgroupplot[xtick = {0, 2, 4, 6, 8, 10},
	xticklabels = {\small$0$, \small$2$,\small $4$, \small $6$, \small $8$, \small $10$},
x grid style={white!80.0!black},
xmajorgrids,
xmin=0.0, xmax=10,
ytick = {0.0, 0.7},
yticklabels = {\small $0$, \small $0.7$},
y grid style={white!80.0!black},
ylabel={$[\text{V}]$},
ymajorgrids,
ymin=-0.2, ymax=0.9,
]
\addplot [thick, black, forget plot] table [x=T, y=X3, col sep=comma]{circuit_simulationData.tex};
\addplot [thick, black, dashed, forget plot] table [x=T, y=X4, col sep=comma]{circuit_simulationData.tex};
\node[] at (axis cs: 0.5, 0.05, 0.16) {$\xi_{3}$};
\node[] at (axis cs: 0.5, 0.7, 0.16) {$\xi_{4}$};

\nextgroupplot[xtick = {0, 2, 4, 6, 8, 10},
	xticklabels = {\small$0$, \small$2$,\small $4$, \small $6$, \small $8$, \small $10$},
x grid style={white!80.0!black},
xmajorgrids,
xmin=0.0, xmax=10,
xlabel={$t \, [\text{s}]$},
ytick = {-15, 0, 10},
yticklabels = {\small $-15$, \small $0$, \small $10$},
y grid style={white!80.0!black},
ylabel={$I[\text{mA}]$},
ymajorgrids,
ymin=-16, ymax=12,
]
\addplot [thick, black, forget plot] table [x=T, y=I, col sep=comma]{circuit_simulationData.tex};
\end{groupplot}

\end{tikzpicture}
	\caption{Time trajectories of states $\xi_{1}, \xi_{2}$ (top) and 
	$\xi_{3}, \xi_{4}$ (middle) 
	of the circuit of Fig.~\ref{fig:circuit:nr}. The extra current input $I$ (bottom)
	confirms the bistable nature of the circuit.}
	\label{fig:sim:1}
\end{figure}
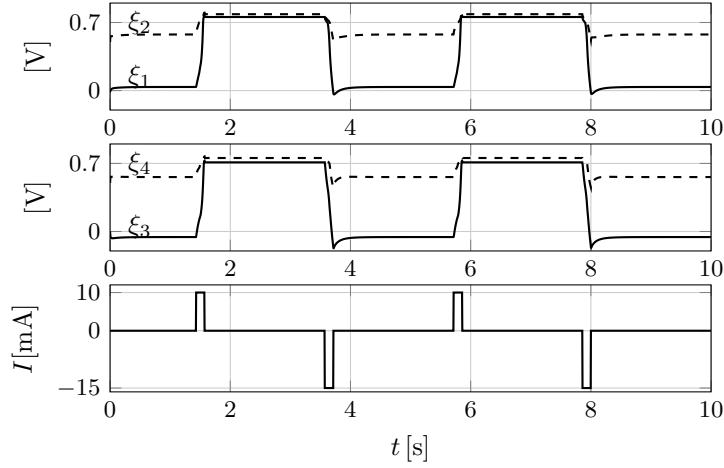

It only rest to show that, with the chosen parameters, the network shows bistability.
To that end, we use the  recent approach of dominant systems developed in~\cite{forni2019,miranda2018b}.
Specifically, with our selection of parameters, the system~\eqref{eq:lcs:1}-\eqref{eq:lcs:2} is strictly $1$-passive from the 
``input'' $z$ to the ``output'' $w$. That is, it satisfies the $\lambda$-parameterized linear matrix inequality
\begin{displaymath}
	\begin{bmatrix}
		A^{\top} P + P A + 2 \lambda P + \varepsilon I & P B_{1} - C^{\top}
		\\
		B_{1}^{\top} P - C & -D - D^{\top}
	\end{bmatrix} \leq 0
\end{displaymath}
for values of $\lambda \in [105, 135]$ and
\begin{displaymath}
P = 
\begin{bmatrix}
	-1.87 & -5.96 & -2.42 & 3.65
	\\
	-5.96 & -5.66 & -5.90 & -0.18
	\\
	-2.42 & -5.90 & -0.86 & 4.42
	\\
	3.65 & -0.18 & 4.42 & 4.67
\end{bmatrix} \times 10^{-3} \;,
\end{displaymath}
which has inertia $(1, 0, 3)$.
Now, since the complementarity condition \eqref{eq:lcs:3} is incrementally passive ($0$-passive in the language of \cite{forni2019}), we conclude that the closed-loop 
\eqref{eq:lcs:1}-\eqref{eq:lcs:3}
is $1$-passive with rate $\lambda > 0$ and therefore $1$-dominant. 
Hence, the asymptotic behavior of the network 
is topologically equivalent to the asymptotic behavior of a $1$-dimensional system.
Further, standard computations show that the trajectories are bounded and there is
one unstable equilibrium, leading us to the desired conclusion.
Fig.~\ref{fig:sim:1} shows the time evolution of the states of the 
network for the choice of parameters described above when an additional current source 
is connected between the terminals $P_{1}-P_{0}$, confirming the desired bistable behavior. 
\section{Classification of two-dimensional LCPs \\ and their stability margins} \label{sec:classification}
In this section, we apply the results on LCP equivalence and stability in order to classify
all possible cone configurations of 2-dimensional LCPs and, therefore, all possible bifurcations of an associated LCS. Our incentive for studying such a
low-dimensional problem in full detail comes from the fact that, in some cases, the singularities of an
LCP can be assessed by studying a lower-dimensional one. 
\begin{exmp}
 Let
 \begin{displaymath}
  M =
   \begin{bmatrix}
    M_{11} & M_{12} \\ 0 & M_{22}
   \end{bmatrix}
  \quad \text{and} \quad 
  q = 
   \begin{bmatrix}
    q_1 \\ q_2
   \end{bmatrix} \;,
 \end{displaymath}
 and consider a triangular $\lcp(M,q)$. Suppose that $M_{22}$ is a $P$-matrix. 
 Then, $x_2$ is uniquely given by 
 \begin{displaymath}
  x_2 = f_{M_{22}}^{-1}(q_2) \;.
 \end{displaymath}
 Moreover, since $M_{22}$ is a $P$-matrix, $f_{M_{22}}$ is regular.
 The problem can be then rewritten as
 \begin{displaymath}
  f_{M_{11}}(x_1) = \bar{q}_1 \;, \quad \bar{q}_1 = M_{12} [-f_{M_{22}}^{-1}(q_2)]^+ + q_1 \;,
 \end{displaymath}
 from where we see that the nonsmooth singularities of the $\lcp(M,q)$ are determined by the
 nonsmooth singularities of the lower-dimensional $\lcp(M_{11},\bar{q}_1)$.
\end{exmp}
Using the characterization of stable LCPs provided by Theorem~\ref{thm:lcp:stable},
we start by enumerating all the unstable matrices. Such enumeration is performed by 
writing down all the conditions leading to weakly degenerate matrices.
For $n = 2$, such classification is relatively easy since, in $\RE^{2 \times 2}$,
weakly degenerate matrices either satisfy $-M_{\cdot, k} 
\in \pos I_{\cdot, k}$, for $k \in \Zint{2}$, or they have degenerate complementary
cones. 
Moreover, it follows from Corollary~\ref{cor:equivalent:diagonal}
that any matrix $N \in \RE^{2 \times 2}$ is LCP-equivalent to the matrix
\begin{equation}
 \label{eq:polar:matrix}
 M (\theta_{1}, \theta_{2}) = 
 \begin{bmatrix}
	r_{1} \cos (\theta_{1} + \pi) & r_{2} \cos (\theta_{2} + \frac{3 \pi}{2})\\
   r_{1} \sin (\theta_{1} + \pi) & r_{2} \sin ( \theta_{2} + \frac{3 \pi}{2})
 \end{bmatrix} \;,
\end{equation}
where $r_{1}, r_{2} \in \{0, 1\}$ and $(\theta_{1}, \theta_{2}) \in [0, 2 \pi) \times 
[0, 2 \pi)$ are angles measured in counterclockwise direction with respect to the rays 
$\pos I_{\cdot, 1}$ and $\pos I_{\cdot, 2}$, respectively.
In what follows, we use the representation~\eqref{eq:polar:matrix} for enumerating all 
$2 \times 2$ unstable matrices. First we consider the class of strongly degenerate matrices, 
for which at least one column is zero. It is characterized by the conditions
\begin{equation} \label{eq:class:0}
\begin{aligned}
	r_{1} r_{2} & = 0 \\
	(\theta_{1}, \theta_{2}) & \in [0, 2\pi) \times [0, 2 \pi) 
\end{aligned} \;.
\end{equation}
The second class of unstable matrices encompasses the strongly degenerate matrices for 
which at least one complementary cone is a subspace of $\RE^{n}$. This class is 
characterized by the conditions
\begin{equation}
	\begin{aligned}
		r_{1} = r_{2} & = 1 \;,
		\\
		(\theta_{1}, \theta_{2}) & \in \Theta^+
	\bigcup \left( \left\{ \frac{3 \pi}{2} \right\} \times [0, 2 \pi]  \right)
	\\
	& \quad \bigcup \left( [0, 2 \pi) \times \left\{ \frac{\pi}{2} \right\} \right) \;,
	\end{aligned} 
	\label{eq:class:1}
\end{equation}
where 
\begin{displaymath}
	\Theta^+ = \left\{ (\theta_1, \theta_{2}) \mid \theta_{1} \in [0, 2\pi), 
	\theta_{2} = \left( \theta_{1} + \frac{\pi}{2} \right) 
\bmod{2 \pi} \right\} \;.
\end{displaymath}
Certainly, matrices satisfying either~\eqref{eq:class:0} or~\eqref{eq:class:1}, 
lie outside of the $R_{0}$-class. 
Finally, we consider the class of unstable $R_{0}$-matrices embracing the remaining degenerate 
and weakly degenerate matrices. This class is characterized by the conditions
\begin{equation}
	\begin{aligned}
		r_{1} = r_{2} & = 1 \;,
		\\
		(\theta_{1}, \theta_{2}) & \in \Theta^{-} 
		\bigcup
		\left( \left\{ \frac{\pi}{2} \right\} \times [0, 2 \pi) \right)
		\\
	& \quad \bigcup \left( [0,  2 \pi)] \times \left\{ \frac{3 \pi}{2} \right\} \right) \;,
	\end{aligned}
	\label{eq:class:2}
\end{equation}
where 
\begin{displaymath}
	\Theta^{-} := \left\{ (\theta_{1}, \theta_{2}) \mid \theta_{1} \in [0, 2\pi), 
	\theta_{2} = \left( \theta_{1} - \frac{\pi}{2} \right) \bmod{2 \pi} \right\} \;.
\end{displaymath} 
The list of unstable normal forms complements the stable normal forms enumerated
in~\cite{castanos2020}. However, the list~\eqref{eq:class:0}-\eqref{eq:class:2} is more 
fundamental, in the sense that representatives of all stable classes can be obtained
as perturbations of conditions \eqref{eq:class:0}-\eqref{eq:class:2}.
\begin{figure}[ht]
	\centering
	\includegraphics{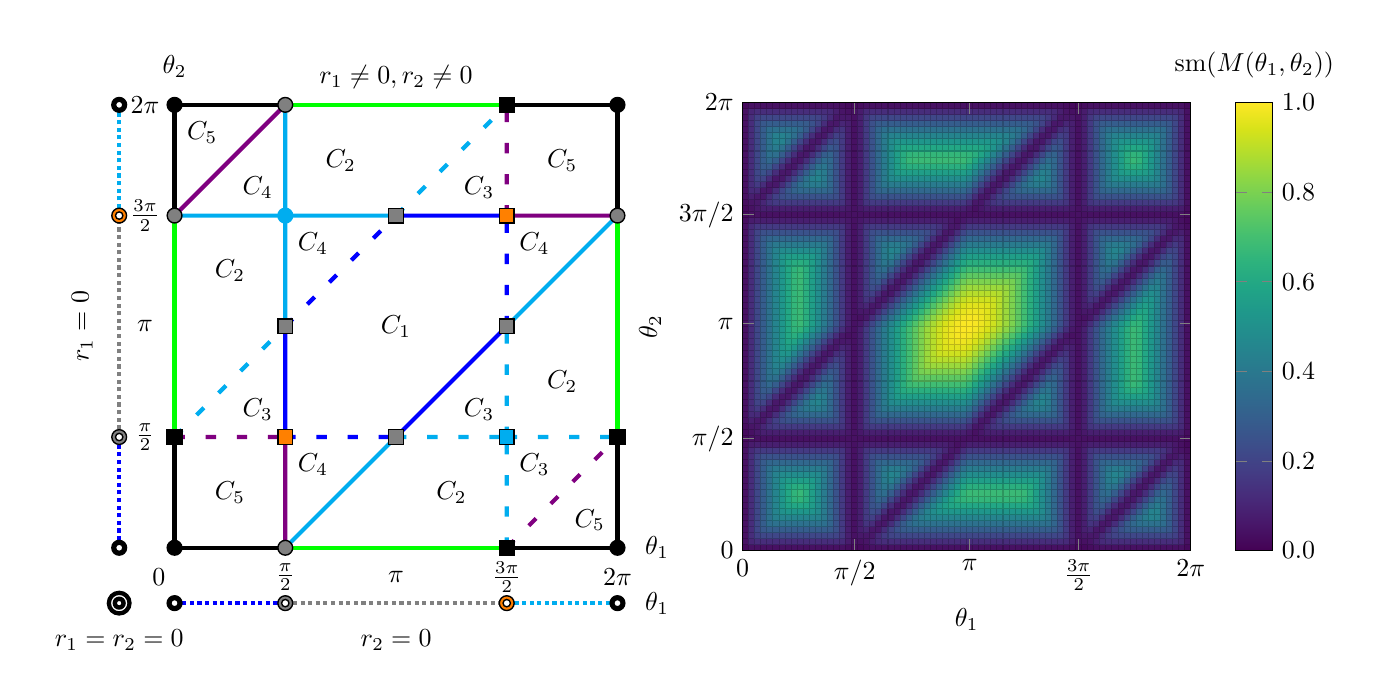}
	\caption{(Left) Classification of $2 \times 2$ matrices~\eqref{eq:polar:matrix}. 
	Line segments and points with the same color, style and 
	shape are LCP-equivalent.
	In the central square, regions of stable matrices ($C_{1}$--$C_{5}$)
	are delimited by the unstable matrices 
	satisfying \eqref{eq:class:0}-\eqref{eq:class:2}.
	(Right) Stability margin \eqref{eq:stab:margin} of 
	$2 \times 2$ matrices in the space of parameters $(\theta_{1}, \theta_{2})$ for the
	case $r_{1} \neq 0$ and $r_{2} \neq 0$. 
	Note that local maxima are closer to the centers of mass of the stable 
	regions.}
	\label{fig:bifDiag}
\end{figure}
\begin{figure}[tb]
	\centering
	\input{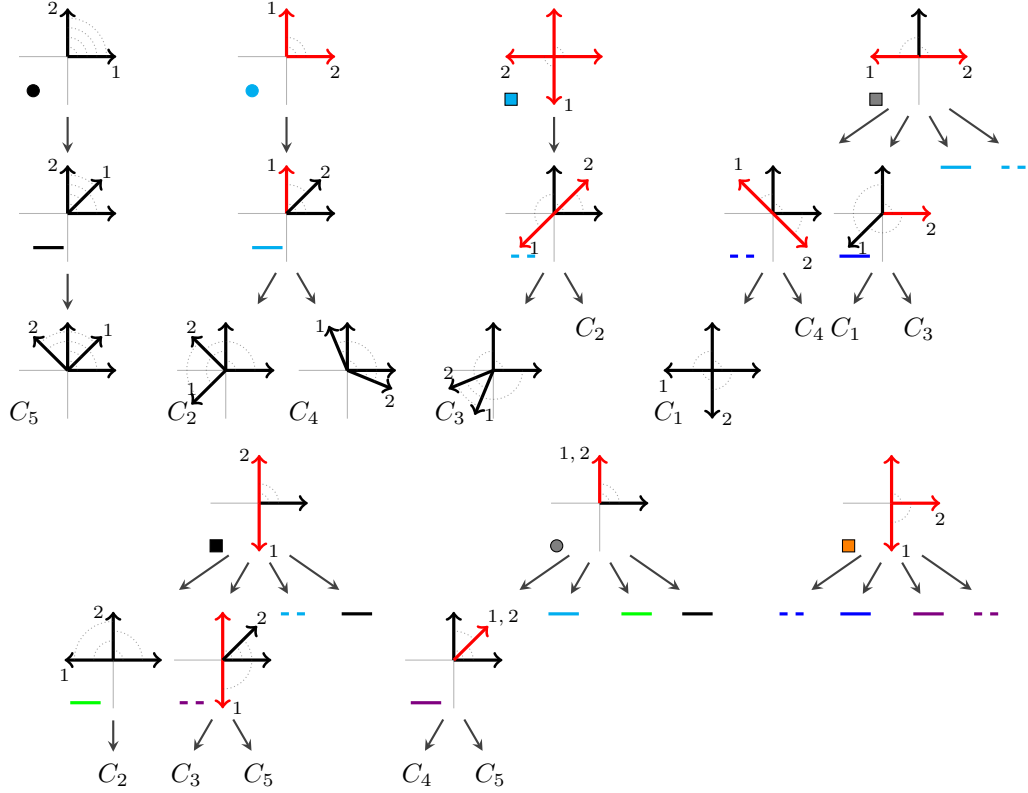}
	\caption{Complementary cone configurations of representative members of the classes
	 depicted in the central square in Fig.~\ref{fig:bifDiag}. Highly-degenerate configurations (circles and 
    squares) break down into less-degenerate configurations (solid and dashed lines), which in turn break down
    into the stable configurations ($C_1-C_5$). Red rays denote degenerate cones and the 
    labels $1$ and $2$ on the rays stand for the rays $\pos (-M_{\cdot, 1})$ and 
    $\pos (-M_{\cdot, 2})$, respectively.}
	\label{fig:unstable:classes}
\end{figure}
Figure~\ref{fig:bifDiag} illustrates the partition of the space of $2 \times 2$ matrices at 
different levels. Line segments and points with the same color, style and shape are LCP-equivalent.
Unstable matrices satisfying~\eqref{eq:class:0},~\eqref{eq:class:1}, and~\eqref{eq:class:2}
are depicted using dotted, dashed, and solid lines, respectively.
The double annular mark at the bottom left corner represents the zero matrix 
($r_{1} = r_{2} = 0$). Perturbations of this matrix will lead to any of the other three cases: a point on
the left vertical line characterizing the cases where $r_{1} = 0$, a point on the bottom horizontal line 
characterizing the cases where $r_{2} = 0$, or a point on the central square for which 
$r_{1} \neq 0$ and $r_{2} \neq 0$.
Connected white regions on the central square correspond to stable matrices. 
It turns out that there are five different 
classes of stable matrices\footnote{In~\cite{castanos2020}, the classes 
$C_2$ and $C_4$ were mistakenly taken to be the same one.} in $\RE^{2 \times 2}$. 
The complementary cones of representative members of each class are shown in 
Fig.~\ref{fig:unstable:classes}.
From Example \ref{examp:sm} we concluded that the stability margin of the matrix in 
\eqref{eq:example:sm} reaches its maximum value at 
\begin{displaymath}
	\varepsilon \in \{-1.37, 0.37, 0.64, 2.37\}\,.
\end{displaymath} 
In terms of the normal form \eqref{eq:polar:matrix} those points correspond to the cases,
\begin{displaymath}
	(\theta_{1}, \theta_{2}) \in \left\{ \left( \frac{\pi}{6}, \frac{11 \pi}{6} \right),
	\left( \frac{11 \pi}{6}, \frac{\pi}{6} \right), \left( \frac{5 \pi}{3}, \frac{\pi}{3} \right), \left( \frac{4 \pi}{3}, \frac{2 \pi}{3} \right)\right\} \,.
\end{displaymath}
Simple computations show that local maxima are attained at the center of mass of 
regions delimiting stable classes (see right-hand side of Fig.~\ref{fig:bifDiag}). 
It is verified that the stability margin indeed characterizes the stable and unstable 
configurations shown in the left-hand side of Fig.~\ref{fig:bifDiag}, thus providing a 
computational tool that can be used for assessing the LCP-stability of larger matrices. 
We conclude this section by pointing out that the classification diagrams in 
Figs.~\ref{fig:bifDiag}-\ref{fig:unstable:classes} allow us to derive some properties and 
relations concerning well-known families of matrices associated to LCPs~\cite{cottle2010}.
For instance, the union of all LCP-stable matrices together with their solid-line 
boundaries match the set of $2 \times 2$ $R_{0}$-matrices, which is open~\cite{doverspike1982}. 
The class $C_{1}$ agrees with the set of $P$-matrices for which there is a unique 
solution to $\lcp(M,q)$ for any $q \in \RE^{2}$, whereas the closure of $C_{1}$ is
the set of $P_{0}$-matrices. The set of $Q$-matrices, for which $\lcp(M,q)$ admits at 
least one solution for any $q \in \RE^{2}$, is the union of the stable classes 
$C_{1}$ and $C_{3}$, together with their shared boundaries 
(blue solid lines in Fig.~\ref{fig:bifDiag}) which is also an open set. 
It is worth to remark that the set of $Q$ matrices is not connected, 
since $M(\theta_{1}, \frac{\pi}{2})$ is not a $Q$-matrix for any 
$\theta_{1} \in [0, 2\pi)$. A matrix $M$ is copositive if $x^{\top} M x \geq 0$ for all
$x \in \RE_{+}^{n}$, it is strictly copositive if $x^{\top} M x > 0$ for all 
$x \in \RE_{+}^{n} \setminus \{0\}$. Strictly copositive matrices are important from a 
algorithmic viewpoint, as for such matrices the convergence of Lemke's algorithm 
towards a solution is guaranteed~\cite{cottle1968}.
\begin{prop} \label{prop:Q:strict:cop}
	Any $2 \times 2$ $Q$-matrix is $LCP$-equivalent to a strictly copositive matrix.
\end{prop}
The fact that strictly copositive matrices are $Q$-matrices holds for the general case in 
$\RE^{n}$ \cite[Theorem 3.8.5]{cottle}. However, in general,
the converse does not hold. That is, not all $Q$-matrices are strictly copositive. 
Thus, the relevance of Proposition \ref{prop:Q:strict:cop} relies on the fact that, 
in the planar case, there is a bijection between the solutions of $\lcp (M_{1},q_{1})$, 
with $M_{1}$ a $Q$-matrix, and the solutions of $\lcp (M_{2}, q_{2})$ for some $M_{2}$ 
strictly copositive. The LCP-equivalence between $Q$ 
and strictly copositive matrices rests an open question in the general case.
\section{Conclusions} \label{sec:conclusions}
Motivated by developing a general non-smooth bifurcation theory, we related LCS
equilibrium bifurcations with LCP singularities and leveraged the rich LCP geometry to
derive a notion of equivalence between LCPs that is akin the classical notion of topological
equivalence at the foundations of smooth bifurcation theory.
Our notion of equivalence allows us to characterize both stable and unstable (from the viewpoint of 
bifurcations) of LCPs and therefore to determine under which condition a model is fragile (i.e., prone to 
lose its bifurcation behavior) or robust (i.e., prone to preserve its bifurcation behavior) under parameter 
perturbations.
We illustrated our theory on a fairly general prototypical negative resistance circuit using resistors and
bipolar transistors. For this circuit, our methods provides constructive conditions to modulate the circuit
behavior through parameter tuning.
We also applied our notion of equivalence to the full characterization of stable and unstable LCPs 
in two dimensions and compute their stability margins.
Future work will consider the extension of our notion of equivalence to allow complementarity problems 
of different dimensions, as well as extensions to other types of complementarity problems
and general (i.e., not purely equilibrium) bifurcations on linear complementarity systems.
\appendix
\section{Proofs}
\begin{proof}[Of Lemma~\ref{lemma:pwl:f:projection}]
	It follows from \eqref{eq:complementary:matrix} that, for $x \in \pos C_{I}(\alpha)$,
	we have $x_{j} \leq 0$ if $j \in \alpha$, whereas $x_{j} \geq 0$ if $j \notin \alpha$. 
   Then, equation~\eqref{eq:f:piecewise:linear} can be written as
	\begin{align*}
		f_{M}(x) &= \sum_{j = 1}^{n} C_{-M}(\alpha)_{\bscdot, j} x_{j}
		 = \sum_{j \notin \alpha} I_{\bscdot, j} x_{j} + \sum_{j \in \alpha} 
		 M_{\bscdot, j} x_{j} \\
		&= \sum_{j = 1}^{n} I_{\bscdot, j} [x_{j}]^{+} -
		 \sum_{j=1}^{n} M_{\bscdot, j} [-x_{j}]^{+} \;,
	\end{align*}
   from which~\eqref{eq:pwl:f:projection} follows directly.
\end{proof}
\begin{proof}[Of Theorem~\ref{thm:equilibria}]
 It follows from Lemma \ref{lemma:pwl:f:projection} that $f_{D}(x) = [x]^{+} - D [-x]^{+}$ and 
 $f_{M}(x) = [x]^{+} - (D - CA^{-1}B)[-x]^{+}$, leading to
 \begin{equation} \label{eq:fM:fD}
  f_{M}(x) = f_{D}(x) + C A^{-1}B[-x]^{+} \;.
 \end{equation}
 Suppose that $F(\xi^\ast, r, s) = 0$ and set $x^\ast=X(\xi^\ast,s)$. By~\eqref{eq:xi to x}, we have
 \begin{equation} \label{eq:xi:fD}
  f_{D}(x^\ast) = C \xi^\ast + E_{2} s
 \end{equation}
 which, when substituted in \eqref{eq:fM:fD}, yields
 \begin{equation} \label{eq:fM:1}
  f_{M}(x^\ast) = C \xi^\ast + E_{2} s + C A^{-1} B [-x^\ast]^{+} \;.
 \end{equation}
 Since $A$ is nonsingular, we can use~\eqref{eq:F def} and rewrite $F(\xi^\ast, r, s) = 0$ as
 \begin{equation} \label{eq:xi:0}
  \xi^\ast = -A^{-1}(B[-x^\ast]^{+} + E_{1}r) \;.
 \end{equation}	
 Substitution of \eqref{eq:xi:0} into \eqref{eq:fM:1} leads us to $f_{M}(x^\ast) = \bar q(r,s)$.
	
 Conversely, suppose that $f_{M}(x^\ast) = \bar{q}(r,s)$. Equation~\eqref{eq:fM:fD} implies that
 \begin{equation} \label{eq:fM:q}
  f_{D}(x^\ast) + C A^{-1}B[-x^\ast]^{+} = E_2 s - CA^{-1}E_1 r \;.
 \end{equation}
 Set $\xi^\ast = \Xi(x^\ast,r)$. It follows from the definition of $\Xi$ that
 \begin{displaymath}
  C\xi^\ast + E_2 s = -CA^{-1}\left( B[-x^\ast]^++E_1 r \right) + E_2 s
 \end{displaymath}
 which, according to~\eqref{eq:fM:q}, gives~\eqref{eq:xi:fD}.
 Finally, by combining~\eqref{eq:xi:fD} and~\eqref{eq:x to xi} we obtain
 the equilibrium condition
 \begin{displaymath}
  \xi^\ast = -A^{-1}\left( B[-f_D^{-1}(C \xi^\ast + E_{2} s )]^+ + E_1 r \right) \;.
 \end{displaymath}
\end{proof}
\begin{proof}[Of Corollary~\ref{cor:bifs}]
By Theorem~\ref{thm:equilibria},
 \begin{displaymath}
  \{x \mid  f_M(x)=\bar q(r,s) \} = \{ X(\xi,s) \mid F(\xi,r,s)=0 \}
 \end{displaymath}
and, similarly,
 \begin{displaymath}
 \{ \xi \mid F(\xi,r,s)=0 \} =	\{ \Xi(x,s) \mid  f_M(x)=\bar q(r,s) \} \;.
\end{displaymath}
Then, the map $(r,s)\mapsto\{x \mid  f_M(x)=\bar q(r,s)\}$ can be expressed as the
composition of the map $(r,s)\mapsto\{ \xi \mid F(\xi,r,s)=0 \}$ and the Lipschitz map
$X(\cdot,s)$, and, vice versa, the map $(r,s)\mapsto\{ \xi \mid F(\xi,r,s)=0 \}$ can be expressed
as the composition of the map $(r,s)\mapsto\{x \mid  f_M(x)=\bar q(r,s)\}$ and the Lipschitz map
$\Xi(\cdot,s)$. Recalling that the composition of Lipschitz-continuous maps is Lipschitz-continuous,
the map $(r,s)\mapsto\{ \xi \mid F(\xi,r,s)=0\}$ is single-valued and Lipschitz-continuous 
in a neighborhood of $(\xi_0,r_0,s_0)$ if, and only if, the map 
$(r,s)\mapsto\{x \mid  f_M(x)=\bar q(r,s)\}$ is single-valued and Lipschitz-continuous in a neighborhood 
$(X(\xi_0,s_0),r_0,s_0)$.
\end{proof}
\begin{proof}[Of Theorem~\ref{thm:equivalence:suf}]
   Since $\psi_*$ is an automorphism on $\Fi_{I}$, for any 
   $\alpha \subseteq \Zint{n}$, there exists a 
   unique $\beta \subseteq \Zint{n}$ such that
	$\pos C_{I}(\alpha)= \psi^{-1}(\pos C_{I}(\beta))$.
   It then follows from~\eqref{eq:f:piecewise:linear} and~\eqref{eq:equiv:f_M:f_N} that
   $\varphi^{-1}(C_{-M}(\alpha) x) = C_{-N}(\beta)\psi(x)$ for $x \in \pos C_{I}(\alpha)$.
   By~\eqref{eq:f:cones} we finally have
   \begin{displaymath}
    \varphi^{-1}(\pos C_{M}(\alpha)) = \pos C_{N}(\beta) \;,
   \end{displaymath}
   which shows that $\varphi$ induces a bijection from $\Ge_M$ to $\Ge_N$.
\end{proof}
\begin{proof}[Of Theorem~\ref{thm:equivalence:nec}]
   The bijectivity of $\varphi_* : \Fi_M \to \Fi_N$ implies that, for each 
   $\alpha \subseteq \Zint{n}$, there exists a unique $\beta \subseteq \Zint{n}$ 
   such that
   \begin{equation} \label{eq:bool:iso}
    \varphi_*(\pos C_{M}(\alpha)) = \pos C_{N}(\beta) \;.
   \end{equation}
   We will explicitly construct a piecewise linear map $\psi:\RE^n \to \RE^n$ 
   and verify~\eqref{eq:equiv:f_M:f_N}. By~\eqref{eq:f:cones} and~\eqref{eq:bool:iso},
   for each $x \in \pos C_{I}(\alpha)$ there	exists an $x' \in \pos C_{I}(\beta)$ such that
	\begin{equation}
		\label{eq:psi:implicit}
		\varphi^{-1}(C_{-M}(\alpha) x) = C_{-N}(\beta) x' \;.
	\end{equation}
   Moreover, $x'$ is unique by the invertibility of $C_{-N}$ (which follows by the
   non-degeneracy of $N$). We define $\psi: x \mapsto x'$ with $x'$ so constructed, so that
   \begin{displaymath}
		\varphi^{-1}(C_{-M}(\alpha) x) = C_{-N}(\beta)\psi(x)
   \end{displaymath}
   for $x \in \pos C_{I}(\alpha)$. This is simply an explicit rewriting of~\eqref{eq:equiv:f_M:f_N}.
	Since $\varphi^{-1}$ is bijective, we can reverse the argument in order to find
   $\psi^{-1}$. The continuity of $\psi$ and $\psi^{-1}$ are simple consequences
   of the continuity of $\varphi$, $\varphi^{-1}$, $f_M$ and $f_N$.
   Finally, note that $\psi$ induces a permutation on $\Ge_I$, and hence an automorphism on $\Fi_I$.
\end{proof}
\begin{proof}[Of Corollary~\ref{cor:equivalent:permutation}]
   Let $\varphi(q') = Pq'$ and $\psi(x) = P^\top x$, and note that $\psi$ induces a Boolean
   automorphism on $\Fi_I$. Also, note that $P [x]^{+} = [P x]^{+}$. The proof then follows
   from~\eqref{eq:pwl:f:projection},
   \opt{twoside}{
   	\begin{multline*}
        f_M(x) = [x]^+ - M[-x]^+ = \\ P\left( [P^\top x]^+ - N[-P^\top x]^+ \right)
 		  = \varphi \circ f_{N}  \circ \psi(x) \;.
 	   \end{multline*}
   }
   \opt{onecolumn}{
   \begin{displaymath}
       f_M(x) = [x]^+ - M[-x]^+ = P\left( [P^\top x]^+ - N[-P^\top x]^+ \right)
		  = \varphi \circ f_{N}  \circ \psi(x) \;.
   \end{displaymath}
   }
\end{proof}
\begin{proof}[Of Corollary~\ref{cor:equivalent:diagonal}]
	For the first equivalence, let $\varphi(q') = Dq'$ and $\psi(x) = D^{-1} x$, and note 
   that $\psi$ is a Boolean automorphism on $\Fi_I$. We also have $D [x]^{+} = [D x]^{+}$.
   As in the previous corollary, $f_M(x) = \varphi \circ f_{N}  \circ \psi(x)$. 
	For the second equivalence, let $\varphi$ be the identity map, $\varphi(q) = q$, and 
	$\psi(x) = C_{-D}(\alpha) x$ for $x \in \pos C_{I}(\alpha)$.
	Both $\varphi$ and $\psi$ are continuous and invertible.
	Hence, $f_{M D}(x) =  
	 C_{-M D}(\alpha) x = C_{-M}(\alpha) C_{-D}(\alpha) x = \varphi \circ f_{M} \circ
   \psi(x)$, for some $\alpha \in \Zint{n}$ and 
   the conclusion follows.
\end{proof}
\begin{proof}[Of Corollary~\ref{cor:ppt:equivalence}]
	If $\beta = \emptyset$ the result holds trivially, so we suppose that
	$\beta \neq \emptyset$. Let $P \in \RE^{n \times n}$ be the permutation 
	matrix such that $P^{\top} \xi = [ \xi_{\beta}^{\top}, \xi_{\beta^{c}}^{\top}]^{\top}$
	and define $\psi(x) = C_{I}(\beta) x$. 
	It follows from \eqref{eq:pwl:f:projection} that
	\begin{align*}
		& P^{\top} ( f_{M} \circ \psi )(x) 
		\\
		& = [P^{\top} C_{I}(\beta) x]^{+} - 
		P^{\top} M P [-P^{\top} C_{I}(\beta) x]^{+}
		\\
		& = 
		\begin{bmatrix}
			- x_{\beta} \\ x_{\beta^{c}}
		\end{bmatrix}^{+} - 
		\begin{bmatrix}
			M_{\beta ,\beta} & M_{\beta, \beta^{c}}
			\\
			M_{\beta^{c}, \beta} & M_{\beta^{c}, \beta^{c}}
		\end{bmatrix}
		\begin{bmatrix}
			x_{\beta} \\ -x_{\beta^{c}}
		\end{bmatrix}^{+}
		\\
		&  = 
		\begin{bmatrix}
				-M_{\beta, \beta} & 0
				\\
				-M_{\beta^{c}, \beta} & I_{\vert \beta^{c} \vert}
			\end{bmatrix}
		\begin{bmatrix}
			x_{\beta} \\ x_{\beta^{c}}
		\end{bmatrix}^{+}
		 -
		\begin{bmatrix}
			-I_{\vert \beta \vert} & M_{\beta, \beta^{c}}
			\\
			0 & M_{\beta^{c}, \beta^{c}}
		\end{bmatrix}
		\begin{bmatrix}
			-x_{\beta} \\ - x_{\beta^{c}}
		\end{bmatrix}^{+}
	\end{align*}
	\begin{align*}
	& = 
	\begin{bmatrix}
		-M_{\beta, \beta} & 0
		\\
		-M_{\beta^{c}, \beta} & I_{\vert \beta^{c} \vert}
	\end{bmatrix} \left( [P^{\top} x]^{+} - P^{\top} N P [- P^{\top} x]^{+} \right)
	\\
	& = \begin{bmatrix}
		-M_{\beta, \beta} & 0
		\\
		-M_{\beta^{c}, \beta} & I_{\vert \beta^{c} \vert}
	\end{bmatrix} P^{\top} f_{N}(x)
\end{align*}
Hence, $f_{N} = \varphi\circ f_{M} \circ \psi$ with
\begin{equation} \label{eq:varphi:pivot}
	\varphi(q') = P
	\begin{bmatrix}
		-M_{\beta, \beta}^{-1} & 0
		\\
		- M_{\beta^{c}, \beta} M_{\beta, \beta}^{-1} & I_{\vert \beta^{c}\vert}
	\end{bmatrix} P^{\top} q' \;.
\end{equation}
Therefore, $N \sim M$.
\end{proof}
\begin{proof}[Of Lemma~\ref{lemma:complementary:partition:M}]
 	 Assume first that $\pos C_{M}(\alpha)$ is a solid cone. Then,
	 there exists $j_{1} \in \Zint{r}$ such that $Q_{M}^{j_{1}} 
	 \subset \pos C_{M}(\alpha)$.
    Let $\mu(E)$ be the Lebesgue measure of $E \subset \RE^{n}$. If
    \begin{displaymath}
	  \mu( \pos C_{M}(\alpha) \setminus Q_{M}^{j_{1}}) = 0 \;,
    \end{displaymath}
    then $\pos C_{M}(\alpha) = \closure Q_{M}^{j_{1}}$, as desired.
    If
    \begin{displaymath}
  	  \mu( \pos C_{M}(\alpha) \setminus Q_{M}^{j_{1}}) > 0 \;,
    \end{displaymath}
    then there exists $j_{2} \in \Zint{r}$, $j_{2} \neq j_{1}$, such that
    $Q_{M}^{j_{2}} \subset \left(\pos C_{M}(\alpha) \setminus Q_{M}^{j_{1}}\right)$.
   As before, if 
   \begin{displaymath}
	   \mu \left(\pos C_{M}(\alpha) \setminus \left( Q_{M}^{j_{1}} 
	   \cup Q_{M}^{j_{2}} \right) \right) = 0 \;,
   \end{displaymath} 
   then $\pos C_{M}(\alpha) = \closure \left( Q_{M}^{j_{1}} \cup 
Q_{M}^{j_{2}} \right)$, as desired. Since $\pos C_{M}(\alpha)$ is strictly pointed
   and $\Pa_{M}$ is finite and the union is dense in $\RE^{n}$ 
   (property \ref{it:cover}) above), the procedure just introduced stops after a 
   finite number of iterations with $J \subset \Zint{r}$ and the result follows. 
   Suppose now that $\pos C_{M}(\alpha)$ is degenerate. Then,
   $\pos C_{M}(\alpha) \subset \bigcup_{j \in J} \boundary Q_{M}^{j} \subset \mathcal{K}(M)$
   for some $J \subset \Zint{r}$.
\end{proof}
\begin{proof}[Of Corollary~\ref{cor:partition:cones}]
	Suppose that $M \sim N$. Then, there is a homeomorphism $\varphi: \RE^{n} \to \RE^{n}$ 
	such that 
   \begin{displaymath}
    \varphi^{-1}(\boundary \pos C_{M}(\alpha)) = \boundary \pos C_{N}(\beta(\alpha))
   \end{displaymath}
	and
   \begin{displaymath}
    \varphi^{-1}(\interior C_{M}(\alpha)) = \interior C_{N}(\beta(\alpha)) \;,
   \end{displaymath}
	for all $\alpha \subset \Zint{n}$ and some 
   \begin{displaymath}
    \beta: \mathcal{P}(\Zint{n}) \to \mathcal{P}(\Zint{n}) \;.
   \end{displaymath}
   Thus, $\varphi^{-1}$ maps connected regions of $\RE^{n} \setminus \mathcal{K}(M)$ into connected regions
   of $\RE^{n} \setminus \mathcal{K}(N)$, thereby inducing a bijection
   between $\Pa_{M}$ and $\Pa_{N}$.
   On the other hand, if $\hat{\varphi}$ induces a bijection $\hat{\varphi}_*$, it follows from 
   Lemma~\ref{lemma:complementary:partition:M} and the continuity of $\hat{\varphi}$, that
   $\hat{\varphi}$ also induces an isomorphism between $\Fi_{M}$ and $\Fi_{N}$.
\end{proof}
\begin{proof}[Of Lemma~\ref{lem:nowhereDense}]
	By definition, a weakly degenerate matrix $M$ belongs to the union of the set of 
	singular matrices and the set
   \opt{twoside}{
 	 \begin{multline} \label{eq:weak:H}
 	 	H : = \left\{ M \in \RE^{n \times n} \mid \text{there exist $k \in \Zint{n}$ and} \right. \\ 
        \left. S \in \mathcal{T}_{k}(M) \text{ such that $-M_{\cdot, k} \in S$} \right\} \;.
 	 \end{multline}
   }
   \opt{onecolumn}{
    \begin{equation} \label{eq:weak:H}
	  	H : = \left\{ M \in \RE^{n \times n} \mid \text{there exist $k \in \Zint{n}$ and
      $S \in \mathcal{T}_{k}(M)$ such that $-M_{\cdot, k} \in S$} \right\} \;.
	 \end{equation}
   }
	It is well know that the set of singular matrices is nowhere dense in 
	$\RE^{n \times n}$. Indeed, the same is true for the set $H$. 
   Consider $\hat{S} \in \RE^{n \times (n-1)}$ such that $\pos \hat{S} \in \mathcal{T}_{k}(M)$.
	It follows that $\det([\hat{S}, -M_{\cdot, k}]) = 0$ and
	$H$ is nowhere dense. To conclude, recall that the finite union
	of nowhere dense sets is again nowhere dense. 
\end{proof}
\begin{proof}[Of Theorem~\ref{thm:lcp:stable}]
	Let us assume first that some complementary cones of $M$ are degenerate. 
	Since the set of matrices with non-zero principal minors is dense in the 
   set of square matrices, there exists a matrix $N$, arbitrarily close to $M$,
   without degenerate complementary cones. A bijection
   $\varphi_* : \mathcal{G}_{M} \to \mathcal{G}_{N}$ necessarily maps a 
   complementary cone with empty interior onto a complementarity cone with
   non-empty interior. Clearly, $\varphi: \RE^{n} \to \RE^{n}$ cannot be 
   invertible and hence cannot be a homeomorphism. Therefore, $M$ is not LCP-stable.
   Assume now that $M \in H$. Then, there is $J \subset \Zint{n}$ such that,
   for each $k \in J$, there exists a matrix $\hat{S} \in \RE^{n \times n-1}$
   satisfying
   \begin{displaymath}
    \pos [\hat{S}, -M_{\cdot, k}]  = \pos \hat{S} = S \in \mathcal{T}_{k}(M) \;.
   \end{displaymath}
   From Lemma \ref{lem:nowhereDense}, there exists $N \in \RE^{n\times n}$ arbitrarily 
   close to $M$ such that 
   $N$ is not weakly degenerate. 
   In other words, $\det[\hat{S}, -N_{\cdot, k}] \neq 0$, so that 
   $\pos [\hat{S}, -N_{\cdot, k}]$ 
   has non-empty interior. Once again, there is no homeomorphism 
   mapping $\pos [\hat{S}, -M_{\cdot, k}]$ into $\pos [\hat{S}, -N_{\cdot, k}]$. 
   It follows that $M$ is not LCP-stable.
   For sufficiency, let us assume that $M$ is not weakly degenerate.
   There exists a neighborhood $U$ of $M$ such that, for
   every $N \in U$, the partitions $\Pa_M$ and $\Pa_N$ have the same number of cells. Indeed, each cell in 
	$\Pa_M$ is a solid cone by construction and, because there are no
	degenerate cones nor $k \in \Zint{n}$ such that $-M_{\cdot, k} \in S$ for
	some $S \in \mathcal{T}_{k}(M)$, the cells remain solid for sufficiently
   small perturbations. Thus, neither new cells appear nor existing cells disappear in $U$.
   Now, there exist refinements $\hat{\Pa}_M$ and $\hat{\Pa}_N$ of the partitions induced
   by $M$ and $N \in U$, respectively, such that each cell is a convex cone with exactly $n$ generators and
   the number of cells is the same for both partitions. There is thus a one-to-one map
   $\eta: \hat{\Pa}_{M} \to \hat{\Pa}_{N}$ such that
   neighbors are preserved, that is, any two cells in $\hat{\Pa}_{M}$ sharing a 
   $p$-dimensional face are mapped under $\eta$ into cells in $\hat{\Pa}_{N}$ sharing also
   a $p$-dimensional face.
   Let $\hat{Q}_{k}(M)$ and $\hat{Q}_{k}(N) = \eta(\hat{Q}_{k}(M))$ be, 
   respectively, matrices that generate the cells of the new partitions 
   $\hat{\Pa}_M$ and  $\hat{\Pa}_N$, where $k \in \Zint{\hat{r}}$ and $\hat{r} = \vert 
   \hat{\Pa}_{M} \vert = \vert \hat{\Pa}_{N} \vert$.
   Define $\hat{\varphi}: \RE^{n} \to \RE^{n}$ as the piecewise linear function
	\begin{equation}
		\label{eq:homeomorphism:partition}
		\hat{\varphi}(q') = \hat{Q}_{k}(M) \hat{Q}_{k}(N)^{-1} q' \; 
		\text{for $q' \in \pos \hat{Q}_{k}(N)$} \;.
	\end{equation}
	Notice that $\hat{\varphi}$ is continuous by construction, since neighbor cells 
	share $n-1$ generators. That is,
	\begin{displaymath}
		\lim_{\substack{q_i' \to q \\ q_i' \in \pos \hat{Q}_{k}(N)}} 
		\hat{\varphi} (q_i') = \lim_{\substack{q_i' \to q \\ q_i' \in 
		\pos \hat{Q}_{s}(N)}} \hat{\varphi} (q_i')
	\end{displaymath}
	for any $q' \in \pos \hat{Q}_{k}(N) \bigcap \pos \hat{Q}_{s}(N)$. 
   Finally, by similar arguments, $\hat{\varphi}^{-1}$ 
   exists and is also continuous. Furthermore,
   $\hat{\varphi}_*: \Pa_M \to \Pa_N$ is a bijection. 
   Therefore, $M$ and $N$ are LCP-equivalent and,
   since $N$ is arbitrary, we conclude that $M$ is LCP-stable.
\end{proof}
\begin{proof}[Of Corollary~\ref{cor:lcp:stable:principal:minor}]
	Note that, if every minor of $M$ is non-zero, 
	then $M$ cannot be weakly degenerate. Indeed, using Laplace's expansion
   we can see that, for any non-empty sets $\alpha, \beta \subseteq \Zint{n}$ such that
	$\vert \alpha \vert = \vert \beta \vert$,
	\begin{equation}
		\label{eq:minor:formula}
		\vert \det M_{\alpha, \beta} \vert = \vert \det [I_{\cdot, \alpha^{c}},
		M_{\cdot, \beta}] \vert \;.
	\end{equation}
	Thus, our assumption on the minors of $M$ implies that any subset of $n$ vectors of 
	\begin{displaymath}
		\{I_{\cdot, 1}, \dots, I_{\cdot, n}, -M_{\cdot, 1}, \dots, -M_{\cdot, n}\}
	\end{displaymath}
	is linearly independent. Consequently, $M$ does not have degenerate cones and
	$-M_{\cdot, k}$ is not in the linear span of $S$ for all 
	$S \in \mathcal{T}_{k}(M)$. That is, $M$ is not weakly degenerate.
\end{proof}
\begin{proof}[Of Proposition~\ref{prop:hysteresis}]
 Let $\hat{\alpha}_{0} = \{1,3\}$. After lengthy but simple computations, we find that the minor 
 $\det \hat{M}_{\hat{\alpha}_{0}, \hat{\alpha}_{0}} > 0$ if, and only if, $\gamma > 0$;
 whereas the remaining principal minors are always positive. Thus, for $\gamma > 0$, $\hat{M}$ is
 a matrix of class $P$ and the solution of the complementarity problem is unique, regardless of $\hat{q}$.
 If, on the other hand, $\gamma < 0$, then the conditions of Corollary~\ref{cor:hysteresis} hold
 and the result follows. Finally, if $\gamma = 0$, then $\hat{M}$ is degenerate and hence LCP-unstable.
\end{proof}
\begin{proof}[Of Proposition~\ref{prop:Q:strict:cop}]
	As mentioned above, the set of $Q$-matrices is the union of three classes, 
	the classes $C_{1}$, $C_{3}$, and their common boundary. Thus, it suffices to 
	show that there is at least one strictly copositive matrix in each class. 
	To that end, let us consider the set, 
	\begin{displaymath}
		\Lambda := \left\{ (\theta_{1}, \theta_{2}) \mid \theta_{1} \in 
			\left(\pi, \frac{3 \pi}{2}\right), \text{ and } \theta_{2}
		\in \left(\frac{\pi}{2}, \pi \right)  \right\} \,.
	\end{displaymath}
	The conclusion follows by noting that, for any 
	$(\theta_{1}, \theta_{2}) \in \Lambda$, $M(\theta_{1}, \theta_{2})$ is strictly 
	copositive and the fact	that $\Lambda$ intersects the three classes of 
	$Q$-matrices (see Fig.~\ref{fig:bifDiag}).
\end{proof}
\bibliographystyle{IEEEtran}
\bibliography{biblio}

\begin{thebibliography}{10}
\providecommand{\url}[1]{#1}
\csname url@samestyle\endcsname
\providecommand{\newblock}{\relax}
\providecommand{\bibinfo}[2]{#2}
\providecommand{\BIBentrySTDinterwordspacing}{\spaceskip=0pt\relax}
\providecommand{\BIBentryALTinterwordstretchfactor}{4}
\providecommand{\BIBentryALTinterwordspacing}{\spaceskip=\fontdimen2\font plus
\BIBentryALTinterwordstretchfactor\fontdimen3\font minus
  \fontdimen4\font\relax}
\providecommand{\BIBforeignlanguage}[2]{{%
\expandafter\ifx\csname l@#1\endcsname\relax
\typeout{** WARNING: IEEEtran.bst: No hyphenation pattern has been}%
\typeout{** loaded for the language `#1'. Using the pattern for}%
\typeout{** the default language instead.}%
\else
\language=\csname l@#1\endcsname
\fi
#2}}
\providecommand{\BIBdecl}{\relax}
\BIBdecl

\bibitem{boydOpt}
S.~Boyd and L.~Vandenberghe, \emph{Convex Optimization}.\hskip 1em plus 0.5em
  minus 0.4em\relax Cambridge, UK: Cambrige University Press, 2006.

\bibitem{acary2011}
V.~Acary, O.~Bonnefon, and B.~Brogliato, \emph{Nonsmooth modeling and
  simulation for switched circuits}, ser. Lecture Notes in Electrical
  Engineering.\hskip 1em plus 0.5em minus 0.4em\relax Springer, 2011.

\bibitem{adly2017}
S.~Adly, \emph{A Variational Approach to Nonsmooth Dynamics: Applications in
  Unilateral Mechanics and Electronics}.\hskip 1em plus 0.5em minus 0.4em\relax
  Springer, 2017.

\bibitem{goeleven2017}
D.~Goeleven, \emph{Complementarity and variational inequalities in
  electronics}.\hskip 1em plus 0.5em minus 0.4em\relax Academis Press, 2017.

\bibitem{brogliato1999}
B.~Brogliato, \emph{Nonsmooth mechanics: models, dynamics and control},
  2nd~ed.\hskip 1em plus 0.5em minus 0.4em\relax London: Springer-Verlag, 1999.

\bibitem{nagurney1993}
A.~Nagurney, \emph{Network economics: A variational inequality approach}, ser.
  Advances in Computational Economics.\hskip 1em plus 0.5em minus 0.4em\relax
  Springer-Science+Business Media, 1999.

\bibitem{goodfellow2016deep}
I.~Goodfellow, Y.~Bengio, and A.~Courville, \emph{Deep learning}.\hskip 1em
  plus 0.5em minus 0.4em\relax MIT press, 2016.

\bibitem{ferris1997}
M.~C. Ferris and J.~S. Pang, ``Engineering and economic applications of
  complementarity problems,'' \emph{SIAM Reviews}, vol.~39, pp. 669 -- 713,
  1997.

\bibitem{murty1988}
K.~G. Murty, \emph{Linear complementarity, linear and nonlinear
  programming}.\hskip 1em plus 0.5em minus 0.4em\relax Berlin: Helderman
  Verlag, 1988.

\bibitem{isac1992}
G.~Isac, \emph{Complementarity Problems}, ser. Lecture Notes in
  Mathematics.\hskip 1em plus 0.5em minus 0.4em\relax Springer-Verlag, 1992.

\bibitem{leenaerts1998}
D.~M. Leenaerts and W.~M.~G. Bokhoven, \emph{Piecewise linear modeling and
  analysis}, ser. Kluwer Academic.\hskip 1em plus 0.5em minus 0.4em\relax New
  York: Springer, 1998.

\bibitem{billups2000}
S.~C. Billups and K.~G. Murty, ``Complementarity problems,'' \emph{Journal of
  Computational and Applied Mathematics}, vol. 124, pp. 303 -- 318, 2000.

\bibitem{facchinei2003}
F.~Facchinei and J.~S. Pang, \emph{Finite-Dimensional Variational Inequalities
  and Complementarity Problems}, ser. Springer series in operations
  research.\hskip 1em plus 0.5em minus 0.4em\relax New York: Springer, 2003,
  vol.~I.

\bibitem{cottle}
R.~W. Cottle, J.~S. Pang, and R.~E. Stone, \emph{The Linear Complementarity
  Problem}, ser. Classics in Applied Mathematics.\hskip 1em plus 0.5em minus
  0.4em\relax Society for Industrial and Applied Mathematics, 2009.

\bibitem{schaft1998}
A.~J. van~der Schaft and J.~M. Schumacher, ``Complementarity modeling of hybrid
  systems,'' \emph{{IEEE} Trans. Autom. Control}, vol.~43, pp. 483 -- 490,
  1998.

\bibitem{heemels2000}
W.~P. M.~H. Heemels, J.~M. Schumacher, and S.~Weiland, ``Linear complementarity
  systems,'' \emph{SIAM Journal on Applied Mathematics}, vol.~60, no.~4, pp.
  1234 -- 1269, 2000.

\bibitem{camlibel2002}
M.~K. {\c{C}}aml{\i}bel, W.~P. M.~H. Heemels, and J.~M. Schumacher, ``On linear
  passive complementarity systems,'' \emph{European Journal of Control},
  vol.~8, no.~3, pp. 220--237, 2002.

\bibitem{brogliato2003}
B.~Brogliato, ``Some perspectives on the analysis and control of
  complementarity systems,'' \emph{{IEEE} Trans. Autom. Control}, vol.~48,
  no.~6, pp. 918 -- 935, 2003.

\bibitem{casey2006piecewise}
R.~Casey, H.~d. Jong, and J.-L. Gouz{\'e}, ``Piecewise-linear models of genetic
  regulatory networks: equilibria and their stability,'' \emph{Journal of
  mathematical biology}, vol.~52, no.~1, pp. 27--56, 2006.

\bibitem{mckean1970nagumo}
H.~McKean~Jr, ``Nagumo's equation,'' \emph{Advances in mathematics}, vol.~4,
  no.~3, pp. 209--223, 1970.

\bibitem{chen2003bifurcation}
G.~Chen, D.~J. Hill, and X.~Yu, \emph{Bifurcation control: theory and
  applications}.\hskip 1em plus 0.5em minus 0.4em\relax Springer Science \&
  Business Media, 2003, vol. 293.

\bibitem{krener2004control}
A.~J. Krener, W.~Kang, and D.~E. Chang, ``Control bifurcations,'' \emph{IEEE
  Transactions on Automatic Control}, vol.~49, no.~8, pp. 1231--1246, 2004.

\bibitem{abed1986local}
E.~H. Abed and J.-H. Fu, ``Local feedback stabilization and bifurcation
  control, {I}. {Hopf} bifurcation,'' \emph{Systems \& Control Letters},
  vol.~7, no.~1, pp. 11--17, 1986.

\bibitem{abed1987local}
------, ``Local feedback stabilization and bifurcation control, {II}.
  stationary bifurcation,'' \emph{Systems \& Control Letters}, vol.~8, no.~5,
  pp. 467--473, 1987.

\bibitem{chen2000bifurcation}
G.~Chen, J.~L. Moiola, and H.~O. Wang, ``Bifurcation control: theories,
  methods, and applications,'' \emph{International Journal of Bifurcation and
  Chaos}, vol.~10, no.~03, pp. 511--548, 2000.

\bibitem{castanos2017implementing}
F.~Casta\~{n}os and A.~Franci, ``Implementing robust neuromodulation in
  neuromorphic circuits,'' \emph{Neurocomputing}, vol. 233, pp. 3--13, 2017.

\bibitem{franci2014realization}
A.~Franci and R.~Sepulchre, ``Realization of nonlinear behaviors from
  organizing centers,'' in \emph{53rd IEEE Conference on Decision and
  Control}.\hskip 1em plus 0.5em minus 0.4em\relax IEEE, 2014, pp. 56--61.

\bibitem{franci2019sensitivity}
A.~Franci, G.~Drion, and R.~Sepulchre, ``The sensitivity function of excitable
  feedback systems,'' in \emph{2019 IEEE 58th Conference on Decision and
  Control (CDC)}.\hskip 1em plus 0.5em minus 0.4em\relax IEEE, 2019, pp.
  4723--4728.

\bibitem{gray2018multiagent}
R.~Gray, A.~Franci, V.~Srivastava, and N.~E. Leonard, ``Multiagent
  decision-making dynamics inspired by honeybees,'' \emph{IEEE Transactions on
  Control of Network Systems}, vol.~5, no.~2, pp. 793--806, 2018.

\bibitem{franci2021analysis}
A.~Franci, A.~Bizyaeva, S.~Park, and N.~E. Leonard, ``Analysis and control of
  agreement and disagreement opinion cascades,'' \emph{Swarm Intelligence}, pp.
  1--36, 2021.

\bibitem{bizyaeva2020general}
A.~Bizyaeva, A.~Franci, and N.~E. Leonard, ``Nonlinear opinion dynamics with
  tunable sensitivity,'' \emph{arXiv preprint arXiv:2009.04332}, 2020.

\bibitem{franci2019model}
A.~Franci, M.~Golubitsky, A.~Bizyaeva, and N.~E. Leonard, ``A model-independent
  theory of consensus and dissensus decision making,'' \emph{arXiv preprint
  arXiv:1909.05765}, 2019.

\bibitem{franci2022breaking}
A.~Franci, M.~Golubitsky, I.~Stewart, A.~Bizyaeva, and N.~Leonard, ``Breaking
  indecision in multi-agent, multi-option dynamics,'' \emph{SIAM Journal on
  Applied Dynamical Systems}, 2023.

\bibitem{Khalil}
H.~K. Khalil, \emph{Nonlinear Systems}.\hskip 1em plus 0.5em minus 0.4em\relax
  Upper Saddle River, New Jersey: Prentice-Hall, 2002.

\bibitem{makarenkov2012dynamics}
O.~Makarenkov and J.~Lamb, ``Dynamics and bifurcations of nonsmooth systems: A
  survey,'' \emph{Physica D: Nonlinear Phenomena}, vol. 241, no.~22, pp.
  1826--1844, 2012.

\bibitem{di2008bifurcations}
M.~Di~Bernardo, C.~Budd, A.~Champneys, P.~Kowalczyk, A.~Nordmark, G.~Tost, and
  P.~Piiroinen, ``Bifurcations in nonsmooth dynamical systems,'' \emph{SIAM
  review}, vol.~50, no.~4, pp. 629--701, 2008.

\bibitem{castanos2020}
F.~Casta\~{n}os, F.~A. Miranda-Villatoro, and A.~Franci, ``A notion of
  equivalence for linear complementarity problems with application to the
  design of non-smooth bifurcations,'' in \emph{21st IFAC World Congress},
  2020.

\bibitem{eaves1981a}
B.~C. Eaves and C.~E. Lemke, ``Equivalence of {LCP} and {PLS},''
  \emph{Mathematics of Operation Research}, vol.~6, no.~4, pp. 475--484, 1981.

\bibitem{golubitsky}
M.~Golubitsky and D.~G. Schaeffer, \emph{Singularities and Groups in
  Bifurcation Theory: Volume I}.\hskip 1em plus 0.5em minus 0.4em\relax
  Springer Science \& Business Media, 1985, vol.~51.

\bibitem{guckenheimer}
J.~Guckenheimer and P.~Holmes, \emph{Nonlinear Oscillations, Dynamical Systes
  and Bifurcations of Vector Fields}.\hskip 1em plus 0.5em minus 0.4em\relax
  Springer-Verlag, 1983.

\bibitem{acary2014}
V.~Acary, H.~de~Jong, and B.~Brogliato, ``Numerical simulation of
  piecewise-linear models of gene regulatory networks using complementarity
  systems,'' \emph{Physica D}, vol. 269, pp. 103--119, 2014.

\bibitem{carroll1995}
T.~L. Carroll, ``A simple circuit for demonstrating regular and synchronized
  chaos,'' \emph{American Journal of Physics}, vol.~63, no.~4, pp. 377--379,
  1995.

\bibitem{eaves1981b}
B.~C. Eaves and C.~E. Lemke, ``On the equivalence of the linear complementarity
  problem and a system of piecewise linear equations: part {II},'' in
  \emph{Homotopy Methods and Global Convergence}, B.~C. Eaves, F.~J. Gould,
  H.-O. Peitgen, and M.~J. Todd, Eds.\hskip 1em plus 0.5em minus 0.4em\relax
  Plenum Publishing Corporation, 1981, pp. 79--90.

\bibitem{miranda2018}
F.~A. Miranda-Villatoro, B.~Brogliato, and F.~Casta\~{n}os, ``Set-valued
  sliding-mode control of uncertain linear systems: continuous and
  discrete-time analysis,'' \emph{SIAM Journal on Control and Optimization},
  vol.~56, no.~3, pp. 1756--1793, 2018.

\bibitem{brogliato2020}
B.~Brogliato and A.~Tanwani, ``Dynamical systems coupled with monotone
  set-valued operators: formalisms, applications, well-posedness, and
  stability,'' \emph{SIAM Review}, vol.~62, no.~1, pp. 3 -- 129, 2020.

\bibitem{papageorgiou1989}
N.~Papageorgiou, ``Differential inclusions with state constraints,''
  \emph{Proceedings of the Edinburgh Mathematical Society}, vol.~32, pp.
  81--98, 1989.

\bibitem{chua1983}
L.~O. Chua, J.~Yu, and Y.~Yu, ``Negative resistance devices: Part {I},''
  \emph{Circuit Theory and Applications}, vol.~11, pp. 161 -- 186, 1983.

\bibitem{getreu1978}
I.~E. Getreu, \emph{Modeling the Bipolar Transistor}.\hskip 1em plus 0.5em
  minus 0.4em\relax Amsterdam, The Netherlands: Elsevier Scientific Publishing
  Company, 1978.

\bibitem{danao1994}
R.~A. Danao, ``Q-matrices and boundedness of solutions to linear
  complementarity problems,'' \emph{Journal of Optimization Theory and
  Applications}, vol.~83, no.~2, pp. 321--332, 1994.

\bibitem{doverspike1982}
R.~D. Doverspike, ``Some perturbation results for the linear complementarity
  problem,'' \emph{Mathematical Programming}, vol.~23, pp. 181 -- 192, 1982.

\bibitem{garcia1981}
C.~B. Garcia, F.~J. Gould, and T.~R. Turnbull, ``Relations between {PL} maps,
  complementary cones, and degree in linear complementarity problems,'' in
  \emph{Homotopy Methods and Global Convergence}, B.~C. Eaves, F.~J. Gould,
  H.-O. Peitgen, and M.~J. Todd, Eds.\hskip 1em plus 0.5em minus 0.4em\relax
  Plenum Publishing Corporation, 1981, pp. 91--144.

\bibitem{howe1981}
R.~Howe and R.~Stone, ``Linear complementarity and the degree of mappings,'' in
  \emph{Homotopy Methods and Global Convergence}, B.~C. Eaves, F.~J. Gould,
  H.-O. Peitgen, and M.~J. Todd, Eds.\hskip 1em plus 0.5em minus 0.4em\relax
  Plenum Publishing Corporation, 1981, pp. 91--144.

\bibitem{gowda1992}
M.~S. Gowda, ``On the continuity of the solution map in linear complementarity
  problems,'' \emph{SIAM Journal on Optimization}, vol.~2, no.~4, pp. 619--634,
  1992.

\bibitem{robinson1980}
S.~M. Robinson, ``Strongly regular generalized equations,'' \emph{Mathematics
  of Operations Research}, vol.~5, no.~1, 1980.

\bibitem{clarke1990}
F.~H. Clarke, \emph{Optimization and nonsmooth analysis}.\hskip 1em plus 0.5em
  minus 0.4em\relax Society for Industrial and Applied Mathematics, 1990.

\bibitem{givant}
S.~Givant and P.~R. Halmos, \emph{Introduction to Boolean Algebras}.\hskip 1em
  plus 0.5em minus 0.4em\relax New York: Springer-Verlag, 2009.

\bibitem{sikorski}
R.~Sikorski, \emph{Boolean Algebras}.\hskip 1em plus 0.5em minus 0.4em\relax
  Berlin, Germany: Springer-Verlag, 1969.

\bibitem{arnold}
V.~I. Arnold, A.~N. Varchenko, and S.~M. Gusein-Zade, \emph{Singularities of
  Differentiable Maps}.\hskip 1em plus 0.5em minus 0.4em\relax Birkh{\"a}user,
  1985, vol.~1.

\bibitem{samelson1958}
H.~Samelson, R.~M. Thrall, and O.~Wesler, ``A partition theorem for {E}uclidean
  $n$-space,'' \emph{Proceedings of the American Mathematical Society}, vol.~9,
  no.~5, pp. 805 -- 807, 1958.

\bibitem{regis2016}
R.~G. Regis, ``On the properties of positive spanning sets and positive
  bases,'' \emph{Optimization and Engineering}, vol.~17, pp. 229 -- 262, 2016.

\bibitem{forni2019}
F.~Forni and R.~Sepulchre, ``Differential dissipativity theory for dominance
  analysis,'' \emph{{IEEE} Trans. Autom. Control}, vol.~64, pp. 2340 -- 2351,
  2019.

\bibitem{miranda2018b}
F.~A. Miranda-Villatoro, F.~Forni, and R.~Sepulchre, ``Dominance analysis of
  linear complementarity systems,'' in \emph{23rd International Symposium on
  Mathematical Theory of Networks and Systems}, 2018.

\bibitem{cottle2010}
R.~W. Cottle, ``A field guide to the matrix classes found in the literature of
  the linear complementarity problem.'' \emph{Journal of Global Optimization},
  vol.~46, pp. 571 -- 580, 2010.

\bibitem{cottle1968}
R.~W. Cottle and G.~B. Dantzig, ``Complementary pivot theory of mathematical
  programming,'' \emph{Linear Algebra and its Applications}, vol.~1, pp.
  103--125, 1968.

\end{thebibliography}
\end{document}